\documentclass{ws-rv961x669}
\usepackage{subfigure}     
\usepackage{ws-rv-thm}     
\usepackage{ws-rv-van}     

\newcommand{\sub}[1]{\ensuremath{_{#1}}}
\renewcommand{\deg}{\ensuremath{^\circ}}
\newcommand{\mtrx}[2]{{\left[\negthickspace\begin{array}{#1}#2\end{array}\negthickspace\right]}}
\newcommand{\vecb}[1]{\ensuremath{\mathbf{#1}}}

\begin{document}



\chapter[The Measurement of Polarization in Radio Astronomy]
{The Measurement of Polarization in Radio Astronomy}
\label{chap:polarization}
\author[]{Timothy Robishaw}
\address{National Research Council Canada, Herzberg Astronomy and
  Astrophysics Programs, Dominion Radio Astrophysical Observatory, PO Box
  248, Penticton, BC V2A 6J9, Canada,\\ 
tim.robishaw+drao@gmail.com}

\author[T. Robishaw \& C. Heiles]{Carl Heiles}
\address{Department of Astronomy, University of California, Berkeley, CA 94720-3411, USA,\\
heiles@astro.berkeley.edu}



\begin{abstract}

Modern dual-polarization receivers allow a radio telescope to characterize
the full polarization state of incoming insterstellar radio waves.  Many
astronomers incorrectly consider a polarimeter to be the ``backend'' of the
telescope.  We go to lengths to dissuade the reader of this concept: the
backend is the least complicated component of the radio telescope when it
comes to measuring polarization.  The feed, telescope structure, dish
surface, coaxial cables, optical fibers, and electronics can each alter the
polarization state of the received astronomical signal.  We begin with an
overview of polarized radiation, introducing Jones and Stokes vectors, and
then discuss construction of digitized pseudo-Stokes vectors from the
outputs of modern correlators.  We describe the measurement and calibration
process for polarization observations and illustrate how instrumental
polarization can affect a measurement.  Finally, we draw attention to the
confusion generated by various polarization conventions and highlight the
need for observers to state all adopted conventions when reporting
polarization results.

\end{abstract}

\body


\section{Introduction}
\label{rh:sec:intro}

Astronomy involves the reception of light from objects beyond the Earth.
Light from these distant objects can arrive at a telescope with its
electric field having some preferred orientation or rotation.  This
tendency is known as polarization.  Most astronomers are happy to just
measure the intensity of light from distant sources, but radio astronomers
can easily measure the full polarization state of the radio waves they
collect.  Sadly, many astronomers consider polarimetry an esoteric
specialty that's not worth their effort.  The aim of this review is to
offer a clear description of the fundamentals of measuring polarization in
radio astronomy.

At radio wavelengths, we find a number of processes that can produce
polarized radiation:\footnote{We highly recommend Ref.~\refcite{trippe14}
  for a gentle and clear introduction to the general characteristics of
  polarized light and the physical processes that produce polarized
  astronomical radiation.} linearly polarized blackbody emission from the
solid surfaces of planets and moons\citep{heilesd63}; linearly polarized
thermal emission from dust grains aligned with a magnetic field;
synchrotron/cyclotron radiation emitted (or absorbed) by
relativistic/non-relativistic electrons gyrating around magnetic field
lines and producing linearly polarized light; Zeeman splitting of spectral
lines emitted or absorbed in a region threaded by a magnetic field,
producing elliptically polarized light; the Goldreich-Kyalfis effect,
producing linear polarization via scattering of anisotropic spectral-line
radiation by atoms or molecules in a magnetic field; Thomson scattering and
gravitational waves producing linear polarization in the cosmic microwave
background.  Radio sources that show some signs of polarization include our
Sun, planets and moons in the solar system, pulsars, gas clouds in the
interstellar medium, circumstellar disks, masers, synchrotron emission from
galaxies, quasars, jets, and the cosmic microwave background.  Most of
these sources have low fractional polarization (pulsars, solid
surfaces, cyclotron/synchrotron emission, and masers being notable
exceptions, with fractional polarizations up to 100\%).

The polarization of a radio wave can be affected as it travels through
interstellar space.  Faraday rotation causes the polarization angle of a
linearly polarized wave to rotate (by an amount $\propto \lambda^2$) when
the wave traverses an ionized medium threaded by a magnetic field having a
component aligned with the direction of propagation.  The Earth's
ionosphere produces Faraday rotation that must be corrected for; this is a
complicated task for interferometers with intercontinental baselines.

Radio waves then interact with the antenna---typically a dish of some
sort---where they are reflected and brought to a focus.  At the focus the
radio waves in free space are coupled to an antenna, known as the feed.
The feed probes the electric field in an orthogonal basis, typically
orthogonal linear polarizations (which we call X and Y) or left-hand and
right-hand circular polarizations (LCP and RCP).  In this paper, we will
always use the IEEE definition of RCP and LCP (more of this in
\sref{rh:sec:conventions.circular}), for which a receiver would see the
electric vector of incoming radiation rotate counterclockwise and
clockwise, respectively, with time.

From this point forward, the signals are amplified and encounter a large
number of electronic components that change the voltage gain (a complex number;
\sref{rh:sec:mueller}).  In addition, differences in cable length (e.g.,
from the telescope to the backend system) produce a differential phase change
that is proportional to frequency (\sref{rh:sec:cables}), and bandpass
filters incur phase delays (\sref{rh:sec:filters}). Finally, the voltages
are sampled, digitized, correlated, Fourier transformed, and stored
(\sref{rh:sec:correlator}).

In this chapter we discuss how various components of a single-dish radio
telescope system create instrumental polarization and how one corrects
or copes with this.\footnote{If one is interested in the details of
  polarization in interferometers, we refer you to
  Refs.~\refcite{thompsonms01} \& \refcite{smirnov11a}.}

There are some very comprehensive reviews of radioastronomical
polarimetry in the literature\citet{stratenmjr10,straten04,hamakerbs96};
many of them are highly mathematical, employing elegant representations of
polarization and invoking such tricks as Lorentz boosts.  The aficionado
should take the time to understand these papers, and those with a
theoretical bent will really appreciate them, but the polarization
newcomer is likely to be scared away.  It's our opinion that
spectropolarimetrists should be doing more to convince observers to use
this tool rather than obfuscating the methods with complex mathematical
representations.

We begin in \sref{rh:sec:basics} by discussing the basic mathematical
framework of polarization and how polarization is described by electric
fields and, alternatively, by Stokes parameters.  In
\sref{rh:sec:correlator}, we discuss how we digitally create the self-
and cross-products that are necessary for polarization measurement. In
\sref{rh:sec:measurement} we discuss how to create calibrated Stokes
parameters from the digitally created products, including a thorough
accounting of all the processes and components that change the polarization
state of an incoming astronomical radio wave between the feed and the
backend. The off-axis polarization response of a telescope is then
considered in \sref{rh:sec:offaxisinstrumental}. Finally, in
\sref{rh:sec:conventions} we emphasize the important and necessary role
played by polarization conventions---and the unfortunate tendency of
astronomers to ignore those conventions.



\section{Polarization: The Basics}
\label{rh:sec:basics}

\subsection{The Description of Polarization by Electric Fields}

The polarization of a radio wave is defined by the motion of its electric
field vector as a function of time within a plane perpendicular to the
direction of propagation.  That plane is known as the plane of polarization
and the general shape that the electric field traces with time is an
ellipse.  We can quantify this polarization ellipse in terms of any
orthonormal basis in the plane of polarization; in radio astronomy, we
encounter two---the standard Cartesian linear basis and a basis of
circularly rotating unit vectors of opposite handedness.

The electric field vector of a monochromatic light wave travelling along the
$+\hat{z}$ direction can be written in terms of both a linear and circular set
of orthonormal bases:
\begin{equation}
\label{rh:eq:efield}
\begin{split}
\vecb{E}(z,t) = \vecb{E\sub{0}} e^{i(2\pi\nu t - kz)} 
&= \left(\mathcal{E}\sub{x}\vecb{\hat{x}} +
\mathcal{E}\sub{y}\vecb{\hat{y}}\right) e^{i(2\pi\nu t - kz)}\\ 
&= \left(\mathcal{E}\sub{R}\vecb{\hat{R}} +
\mathcal{E}\sub{L}\vecb{\hat{L}}\right) e^{i(2\pi\nu t - kz)} \,,
\end{split}
\end{equation}
where $\vecb{\hat{R}} = (\vecb{\hat{x}} - i\vecb{\hat{y}})/\sqrt{2}$ and
$\vecb{\hat{L}} = (\vecb{\hat{x}} + i\vecb{\hat{y}})/\sqrt{2}$ are the unit
vectors of IEEE RCP and LCP.  As seen from an observer somewhere at
$z>0$ and looking back towards the origin, IEEE RCP is seen to rotate
counterclockwise with time and IEEE LCP clockwise.

We can write $\vecb{E\sub{0}}$ as a Jones vector\citep{jones47} in either
of the bases:
\begin{equation}
\label{rh:eq:full.jones}
\vecb{E\sub{0}} = \mtrx{c}{\mathcal{E}\sub{x} \\ \mathcal{E}\sub{y}} =
\mtrx{c}{ E\sub{0x} e^{i\phi\sub{x}} \\ E\sub{0y} e^{i\phi\sub{y}}}
\ \ {\rm or}\ \ \vecb{E\sub{0}} = \mtrx{c}{\mathcal{E}\sub{R} \\ \mathcal{E}\sub{L}} =
\mtrx{c}{ E\sub{0R} e^{i\phi\sub{R}} \\ E\sub{0L} e^{i\phi\sub{L}}}\,.
\end{equation}
At a given position $z$ along the direction of propagation (let's take
$z=0$ for simplicity), the tip of the electric field vector $\vecb{E}$ will
trace out an ellipse in time with orthogonal components given in the linear
basis by:
\begin{equation}
  \label{rh:eq:electric.field.components.time}
    E\sub{x}(t) = E\sub{0x} e^{i(2\pi\nu t+\phi\sub{x})}\,,\ \ \,\, E\sub{y}(t) = E\sub{0y} e^{i(2\pi\nu t+\phi\sub{y})}\,, \\
\end{equation}
or in the circular basis by:
\begin{equation}
  \label{rh:eq:electric.field.components.time.circular}
    E\sub{R}(t) = E\sub{0R} e^{i(2\pi\nu t+\phi\sub{R})}\,,\ \ E\sub{L}(t) = E\sub{0L} e^{i(2\pi\nu t+\phi\sub{L})}\,.
\end{equation}
These components define the previously mentioned polarization ellipse.
Many treatments of polarization ignore the absolute phase (which must not
be ignored when using an interferometer!) and define the relative phase as
$\Delta\phi \equiv \phi\sub{y} - \phi\sub{x}$.

The major axis of the polarization ellipse will be oriented at an angle
$\chi$ with respect to the $x$ axis (see
\fref{rh:fig:stokes.def}\textit{a}) where
\begin{equation}
  \label{rh:eq:polarization.ellipse.chi}
  \tan 2\chi =
  \frac{2E\sub{0x}E\sub{0y}\cos\left(\phi\sub{y}-\phi\sub{x}\right)}{E\sub{0x}^{2}-E\sub{0y}^{2}}
  = \tan \left(\phi\sub{R}-\phi\sub{L}\right)\ ;
  \hspace{3ex} 0\deg \leq \chi \leq 180\deg \,.
\end{equation}


\subsection{The Description of Polarization by Stokes Parameters}

\begin{figure}[!t]
  \centering
  \includegraphics[width=\textwidth]{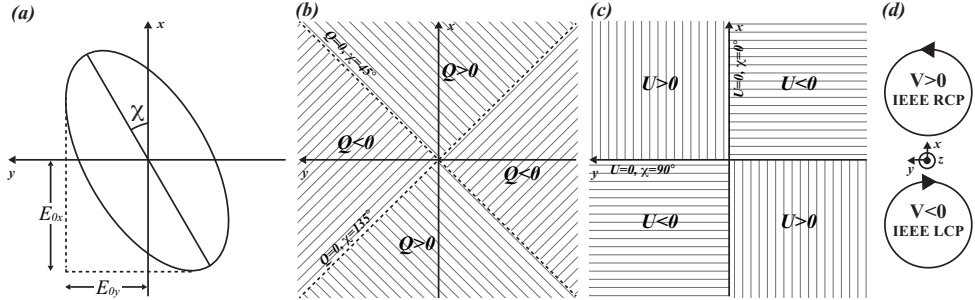}
  \vspace{-2ex}
  \caption{(\textit{a}) The polarization ellipse.  For a radio wave
    travelling along the $z$ axis (out of the page), the electric field
    will trace out an ellipse in the $xy$ plane with time at a given
    position $z$.  The azimuth of the major axis of the ellipse relative to
    the $x$ axis, $\chi$, is known as the \textit{polarization angle}.  IAU
    convention (see \sref{rh:sec:linear.polarization.convention}) aligns
    the $x$ axis toward north on the sky. (\textit{b})-(\textit{c})
    Representations of the sign for Stokes $Q$ and $U$, respectively, given
    the polarization angle of the major axis of the ellipse. (\textit{d})
    Representations of the sign of Stokes $V$ using IEEE and IAU
    conventions (see \sref{rh:sec:conventions.circular}).
\label{rh:fig:stokes.def}}
\end{figure}

Astronomical radio signals are, in general, partially polarized.  The
polarization ellipse and Jones matrices cannot help us quantify partially
polarized radiation.  For this, we use the Stokes parameters.  The Stokes
parameters are most often denoted as $I$, $Q$, $U$, and $V$ in astronomical
measurements and, because they are conveniently manipulated by matrix
algebra, are often written as the Stokes vector,\footnote{While matrices
  are often represented by a bold font, here we have introduced the
  notation $\underline{A}$ to represent a $1\times4$ column matrix---known
  as a \textit{vector} in the parlance of linear algebra---to differentiate
  from a physical vector ${\bf A}$, e.g., the electric field. (The Stokes
  vector comprises the Stokes parameters, which do not represent an
  orthonormal basis: Stokes $I$ can be a linear combination of Stokes $Q$,
  $U$, and $V$.) We later use the notation $\underline{\underline{A}}$ to
  represent a square $4\times4$ matrix.}
\begin{eqnarray} \label{rh:eq:s_def}
\underline{S} = \left[ 
\begin{array}{c} 
    S_0 \\ S_1 \\ S_2 \\ S_3 \\
\end{array} 
\; \right] \; \equiv \left[
\begin{array}{c} 
    I \\ Q \\ U \\ V \\
\end{array} 
\; \right] \; ,
\end{eqnarray} 
where the Stokes parameters are defined\citet{shurcliff62,kligerlr90} in
terms of the intensities of orthogonal polarization forms
$(I\sub{0\deg},I\sub{90\deg})$, $(I\sub{+45\deg},I\sub{-45\deg})$,
and $(I\sub{\rm RCP},I\sub{\rm LCP})$:
\begin{enumerate}

\item Stokes $I$ is the total intensity.  It is the sum of the intensities
  of any two orthogonal polarization components and does not store any
  polarization information.
  $I \equiv I\sub{\rm tot} \equiv I\sub{0\deg} + I\sub{90\deg} \equiv
  I\sub{+45\deg} + I\sub{-45\deg} \equiv I\sub{\rm RCP} + I\sub{\rm
    LCP}$.\footnote{Here we follow Ref.~\refcite{kligerlr90} in using each
    subscripted $I$ to represent intensities of a given polarization form.
    It might appear recursive to then also define the first Stokes
    parameter as $I$, but this is just a notational convention and the
    reader might wish to think of Stokes $I$ as always having an implicit
    ``tot'' subscript to clarify that it represents the \textit{total} of
    intensities in any one pair of orthogonal polarization states.}

\item Stokes $Q$ is the difference in intensities between horizontal and
  vertical linearly polarized components and is a measure of the tendency
  of the radio wave to prefer the horizontal direction. If $Q>0$ there is
  an excess of polarized radiation along the horizontal, while for $Q<0$,
  there is a vertical excess (\fref{rh:fig:stokes.def}\textit{b}).
  $Q \equiv I\sub{0\deg} - I\sub{90\deg}$.

\item Stokes $U$ is the difference in intensities between linearly polarized
components at $+45\deg$ and $-45\deg$ and represents the preference of the
light to be aligned at $+45\deg$, with $U<0$ meaning an excess in linear
polarization at an angle $-45^{\circ}$ to the horizontal (\fref{rh:fig:stokes.def}\textit{c}). $U \equiv
I\sub{+45\deg} - I\sub{-45\deg}$.

\item Stokes $V$ is the difference between the intensities of the RCP and
  LCP components and describes the preference for the light to be RCP.  For
  positive Stokes $V$, there is an excess of RCP over LCP when using the
  IEEE and IAU conventions (see \sref{rh:sec:conventions.circular};
  (\fref{rh:fig:stokes.def}\textit{d})). $V \equiv I\sub{\rm RCP} -
  I\sub{\rm LCP}$.

\end{enumerate}
It's important to note that these are \textit{definitions}. Stokes
himself\citep{stokes52} used the notation $\{A,B,C,D\}$ a century before
Chandrasekhar\citep{chandrasekhar47} settled on $\{I,Q,U,V\}$, the latter
three letters of which were assigned with no motivation. Given
Chandrasekhar's convention, there still remains room for ambiguity and
confusion: for example, $Q$ could have been defined as
$I\sub{90\deg} - I\sub{0\deg}$, and $V$ could have been defined as
I\sub{\rm LCP} - I\sub{\rm RCP} (and often is! See
\sref{rh:sec:conventions}).

The degree of polarization, or fractional polarization, is the ratio of the
intensity of the polarized emission to the total intensity:
\begin{equation} \label{rh:eq:polfrac}
p = \frac{I\sub{\rm pol}}{I\sub{\rm tot}} = \frac{\sqrt{Q^2 + U^2 + V^2}}{I} ;
\hspace{0.3in} 0 \leq p \leq 1\,.
\end{equation}
We can also form fractional linear polarization
\begin{equation}
p\sub{\rm lin} = \frac{\sqrt{Q^2 + U^2}}{I} ;
\hspace{0.3in} 0 \leq p\sub{\rm lin} \leq 1\,,
\end{equation}
and fractional circular polarization
\begin{equation}
p\sub{\rm cir} = \frac{V}{I} ;
\hspace{0.3in} -1 \leq p\sub{\rm cir} \leq 1\,.
\end{equation}
When combining (or spatially smoothing) polarized signals, one must combine
(or smooth) Stokes parameters, not fractional polarizations, linearly
polarized intensities, or polarization angles.\citet{heiles02}

\subsection{Stokes Parameters Expressed in Terms of Electric Fields}

We can also write the Stokes parameters in terms of the time-averaged self-
and cross-products of the electric field components as
\begin{subequations}
  \label{rh:eq:stokes.orthogonal.cartesian}
  \begin{alignat}{3}
    I &\equiv \phantom{-i(}\left<E\sub{x}\overline{E\sub{x}}\right> + \left<E\sub{y}\overline{E\sub{y}}\right> &&\equiv \phantom{-i(}\left<E\sub{R}\overline{E\sub{R}}\right> + \left<E\sub{L}\overline{E\sub{L}}\right>\,, \\
    Q &\equiv \phantom{-i(}\left<E\sub{x}\overline{E\sub{x}}\right> - \left<E\sub{y}\overline{E\sub{y}}\right> &&\equiv \phantom{-i(}\left<E\sub{R}\overline{E\sub{L}}\right> + \left<\overline{E\sub{R}}E\sub{L}\right>\,, \\
    U &\equiv \phantom{-i(}\left<E\sub{x}\overline{E\sub{y}}\right> + \left<\overline{E\sub{x}}E\sub{y}\right> &&\equiv -i\left(\left<E\sub{R}\overline{E\sub{L}}\right> - \left<\overline{E\sub{R}}E\sub{L}\right>\right)\,, \\
    V &\equiv -i\left(\left<E\sub{x}\overline{E\sub{y}}\right> - \left<\overline{E\sub{x}}E\sub{y}\right>\right) &&\equiv \phantom{-i(}\left<E\sub{R}\overline{E\sub{R}}\right> - \left<E\sub{L}\overline{E\sub{L}}\right>\,,
  \end{alignat}
\end{subequations}
where the angle brackets denote a time average of the electric
field,\footnote{This is necessary because the signal being received is
  being treated as quasi-monochromatic.  Such light will not trace out an
  ellipse with time, but the ellipse can be recovered if the products are
  averaged over a time long relative to the period of the radio wave. Even
  for a very fast correlator that could accumulate only 100~ms of data,
  there will be millions of wave periods per integration at radio
  frequencies, which is plenty long to uncover the polarization properties
  of the astronomical radiation.} and the overbar denotes complex
conjugation.\footnote{Textbooks covering polarization tend to denote
  complex conjugation as $A^*$. Many authors reverse terms in some of the
  difference equations because they've either used the physics convention
  for Stokes $V$ as IEEE ${\rm LCP}-{\rm RCP}$ or they've defined the
  exponential propagation argument of the $E$ field as the negative of the
  IEEE convention that we've adopted in \eref{rh:eq:efield}. Finally,
  there is an understood constant on the RHS of each equation
  accounting for the conversion of the square of the $E$ field to a
  temperature or flux density.} 
By substituting
\eref{rh:eq:electric.field.components.time} and \eref{rh:eq:electric.field.components.time.circular}
into \eref{rh:eq:stokes.orthogonal.cartesian}, we derive the more
commonly found representation\footnote{Optics, radiation, and astronomy
  texts usually provide this set of Stokes parameters, and will often
  include their representation as a function of the polarization ellipse
  parameters. The correlation representation of
  \eref{rh:eq:stokes.orthogonal.cartesian} is not widely presented.}
of the Stokes parameters:
\begin{subequations}
  \label{rh:eq:stokes.ellipse.parameters}
  \begin{alignat}{3}
    I &=  \langle E\sub{0x}^2 \rangle + \langle E\sub{0y}^2\rangle &&= \langle E\sub{0R}^2\rangle + \langle E\sub{0L}^2\rangle \,, \\
    Q &= \langle E\sub{0x}^2\rangle - \langle E\sub{0y}^2\rangle   &&=  2 \langle E\sub{0R} E\sub{0L}\rangle \cos \left(\phi\sub{R} - \phi\sub{L}\right)\,, \\
    U &= \phantom{-}2 \langle E\sub{0x} E\sub{0y}\rangle \cos \left(\phi\sub{y} - \phi\sub{x}\right) &&=  2 \langle E\sub{0R} E\sub{0L}\rangle \sin \left(\phi\sub{R} - \phi\sub{L}\right)\,, \\
    V &= -2 \langle E\sub{0x} E\sub{0y}\rangle \sin \left(\phi\sub{y} - \phi\sub{x}\right) &&= \langle E\sub{0R}^2\rangle -  \langle E\sub{0L}^2\rangle \,.
  \end{alignat}
\end{subequations}
\noindent From \eref{rh:eq:stokes.ellipse.parameters} and
\eref{rh:eq:polarization.ellipse.chi}, it can be seen that the angle that
the polarization ellipse makes with the horizontal (i.e., $x$ axis) can be
expressed by
\begin{equation}
  \chi = \frac{1}{2}\tan^{-1}\left(\frac{U}{Q}\right)\,; \hspace{3ex} 0\deg \leq \chi \leq 180\deg\,,
\end{equation}
where $\chi$ is known as the \textit{position angle of linear polarization}
(or, more succinctly, the \emph{polarization angle}) and has a total range
of 180, not 360, degrees. Therefore, $\chi$ has an \emph{orientation}, not
a \emph{direction}.  Line segments are commonly used to represent the
amplitude and orientation of linear polarization on the plane of the sky.
The astronomical community regularly refers to such a line segment as
a \textit{polarization vector} even though a vector has a direction.  We
propose the adoption of the term \textit{segtor}.



\section{Measuring Self- and Cross-Products with Digital Methods}
\label{rh:sec:correlator}

Our dual-polarized receiver system has two orthogonal polarizations, which
we denote by $A$ and $B$ because the discussion applies, unchanged, whether
our feed system is native linear, native circular, or something in
between. Having both polarizations allows us to synthesize all the Stokes
parameters from self- and cross-products of the two polarizations using the
digital equivalent of \eref{rh:eq:stokes.orthogonal.cartesian}.

The time-averaged voltage products are derived from digital samples in one
of two ways.  Historically, the XF correlation technique\footnote{The ``X''
  represents correlation and the ``F'' represents a Fourier transform.}
prevailed because of its simpler hardware requirements. With XF, one uses a
correlation spectrometer, which produces time-averaged auto- and
cross-correlation functions (ACFs and CCFs, respectively). These are
Fourier transformed, usually in a general-purpose computer, to produce
power spectra.  Each ACF is computed for $N$ positive lags; negative lags
are unnecessary because autocorrelations are symmetric with respect to
lag. The ACFs are averaged over time and the Fourier transform (FT) of the
resulting average ACF gives the self-power spectrum. Because the ACF is
symmetric with respect to lag, its Fourier transform is real and symmetric
with frequency, so the self-product power spectrum has $N$ independent
channels.  Symbolically, for polarization $A$ we write
\begin{equation} 
AA = {\rm FT} \langle {\rm ACF}(V_A) \rangle \,.
\end{equation}

The cross-correlation of the two polarizations is not symmetric with lag,
so it must be computed both for $N$ positive and $N$ negative lags. Its FT
is complex with Hermitian symmetry, so the cross-power spectrum can be
regarded as consisting of a real and imaginary part, each with $N$
independent channels. Symbolically, for polarizations $A$ and $B$ we write
\begin{equation} 
\begin{split} 
AB = {\rm Re} \left\{ {\rm FT} \left\langle {\rm CCF}(V_AV_B) \right\rangle
  \right\} \,,\\
BA = {\rm Im} \left\{ {\rm FT} \left\langle {\rm CCF}(V_AV_B) \right\rangle
  \right\} \,.
\end{split}
\end{equation}
Thus, for a native-linear feed connected to the inputs of a digital
spectrometer in such a way that $(A,B)=(X,Y)$, the spectrometer will
produce the four spectra $[XX, YY, XY, YX]$. Similarly, for a
native-circular feed with $(A,B)=(R,L)$, the spectrometer will output
$[RR, LL, RL, LR]$.

Today, the FX technique is favored because of the heavy computing ability
of FPGAs and GPUs. With FX, each polarization is sampled at rate $t_s$ over
time interval $2T$, providing $2N = \frac{2T}{t_s}$ samples.  This block of
data is Fourier transformed, producing a complex transform of $2N$ channels
with Hermitian symmetry having $N$ positive-frequency and $N$
negative-frequency channels. The self-product power spectrum is this FT
times its complex conjugate, and because of the Hermitian symmetry, it is
real with the $N$ negative- and positive-frequency portions
identical. Thus, it is a power spectrum with $N$ independent
channels. Similarly, one calculates cross-product power spectra by
multiplying the Fourier transforms of the two polarizations with both
possibilities of complex conjugate (\eref{rh:eq:crossp_stokes}). This
produces a complex cross-power spectrum having $2N$ independent channels,
split between negative and positive frequencies. This cross-power spectrum
does not have Hermitian symmetry, so has a real part and an imaginary part,
each with $N$ independent channels. Thus, we have four spectra of length
$N$. Symbolically, for the $V_A$ and $V_B$ self-product spectra we write
\begin{equation} 
AA = \left\langle {\rm FT}(V_A) \overline{{\rm FT}(V_A)}\right\rangle \,, \ \ \ \ BB = \left\langle {\rm FT}(V_B) \overline{{\rm FT}(V_B)}\right\rangle \,.
\end{equation}
\noindent The FX spectrometer will return either the complex cross-product
spectrum
\begin{equation}
\left\langle {\rm FT}(V_A) \overline{{\rm FT}(V_B)} \right\rangle 
\ \ \ \ {\rm or} \ \ \ \
\left\langle \overline{{\rm FT}(V_A)} {\rm FT}(V_B) \right\rangle\,,
\end{equation}
but not both.  Since these are a complex conjugate pair, we can
symbolically represent the real and imaginary parts of these cross-product
spectra as:
\begin{equation} \label{rh:eq:crossp_ab}
\begin{split} 
AB &= {\rm Re} \left\{ \left\langle {\rm FT}(V_A) \overline{{\rm FT}(V_B)}
     \right\rangle \right\} = \phantom{-}{\rm Re} \left\{ \left\langle \overline{{\rm FT}(V_A)} {\rm FT}(V_B) \right\rangle\right\}\,,\\
BA &= {\rm Im} \left\{ \left\langle {\rm FT}(V_A) \overline{{\rm FT}(V_B)}
     \right\rangle \right\} = -{\rm Im} \left\{ \left\langle \overline{{\rm FT}(V_A)} {\rm FT}(V_B) \right\rangle\right\}\,.
\end{split}
\end{equation}
\noindent (Note that ambiguity exists in the sign of the BA term because it
won't be known a priori which of the cross-product spectra an FX
spectrometer will output; this is determined via calibration.) The
real-valued Stokes parameter spectra can then be assembled from the self-
and cross-product spectra following
\eref{rh:eq:stokes.orthogonal.cartesian} as:
\begin{equation} \label{rh:eq:crossp_stokes}
\begin{split} 
 \left[ \,\left\langle {\rm FT}(V_A) \overline{{\rm
      FT}(V_A)} \right\rangle 
+
\left\langle \overline{{\rm FT}(V_B)} {\rm FT}(V_B) \right\rangle \,
\right] &= AA + BB \,, \\
 \left[ \,\left\langle {\rm FT}(V_A) \overline{{\rm
      FT}(V_A)} \right\rangle 
-
\left\langle \overline{{\rm FT}(V_B)} {\rm FT}(V_B) \right\rangle \,
\right] &= AA - BB \,, \\
 \left[ \,\left\langle {\rm FT}(V_A) \overline{{\rm
      FT}(V_B)} \right\rangle 
+
\left\langle \overline{{\rm FT}(V_A)} {\rm FT}(V_B) \right\rangle \,
\right] &= 2\, AB \,, \\
-i \, \left[ \, \left\langle {\rm FT}(V_A) \overline{{\rm FT}(V_B)} \right\rangle
  -
\left\langle \overline{{\rm FT}(V_A)} {\rm FT}(V_B)  \right\rangle \,
\right] &= 2\,BA \,.
\end{split}
\end{equation}

Even after these self- and cross-products have been properly
amplitude-calibrated and combined, they do not provide true Stokes
parameters, because the telescope circuitry introduces cross-coupling
and phase shifts. Thus, they do not provide a true Stokes vector as
defined in \eref{rh:eq:s_def} and
\eref{rh:eq:stokes.orthogonal.cartesian}.  Rather, they provide a {\it
  pseudo}-Stokes vector with four pseudo-Stokes parameters.
In this review, we represent pseudo-Stokes vectors by the
special symbol $\cal S$ (the calligraphic $S$). 

Incorporating all of this, the pseudo-Stokes vector assembled from the
correlator output is
\begin{eqnarray} \label{rh:eq:cor}
\underline{{\cal S}^{\rm cor}} = \left[ 
\begin{array}{c} 
{\cal S}^{\rm cor}_0 \\ {\cal S}^{\rm cor}_1 \\ {\cal S}^{\rm cor}_2 \\ {\cal S}^{\rm cor}_3 \\
\end{array} 
\; \right] = \left[
\begin{array}{c} 
  AA + BB \\
  AA - BB \\
   2\,AB \\
   2\,BA \\ 
\end{array} 
\; \right] \; .
\end{eqnarray} 



\section{The Measurement and Calibration Process} 
\label{rh:sec:measurement}

We've treated everything in our system---from the source's radiation
incident on the Earth to the digital backend output---as a black box.
To convert the resulting pseudo-Stokes vector into a true
Stokes vector for the astronomical source being observed, 
we need to undo the effects of this black box. 

\subsection{Amplitude Calibration}

The digitally produced pseudo-Stokes vector is generated in terms of
arbitrarily scaled numbers derived from the correlator input voltages
$(V_A,V_B)$, which are instrumentally generated from the incoming electric
fields $(E_A,E_B)$. We must convert these arbitrary units to physically
meaningful units (kelvins or janskys), which is done by inserting noise of
known intensity using standard radioastronomical techniques. This is a
standard process that is covered in other chapters of this book, so in the
interest of brevity we will omit further discussion of this topic. Rather,
we assume at this point that all the pseudo-Stokes vector $\underline{{\cal
    S}^{\rm cor}}$ elements are properly calibrated with respect to
amplitude and are brightness temperatures in units of kelvins.\footnote{See Chapter 1 of this volume.}

\subsection{For Illustrative Purposes: A Linearly Polarized Source}

Measuring the polarization of a source means obtaining its four calibrated
Stokes parameters. Here we focus the discussion for illustrative purposes
by considering linearly polarized sources.  Concentrating on linearly
polarized sources is natural, because many polarized radioastronomical
sources have only very small circular polarization; pulsars and masers can
be exceptions.

Purely linearly polarized sources have $V=0$ and are characterized by
the fractional linear polarization $p_{\rm src,lin}$ and the position
angle $\chi_{\rm src}$; these, in turn, are derived from Stokes $(Q,U)$:
\begin{eqnarray} \label{rh:eq:s_src}
\underline{S_{\rm src}} = I_{\rm src} \cdot \left[ 
\begin{array}{c} 
    1 \\ p_{\rm src,lin} \cos 2 \chi_{\rm src} \\ p_{\rm src,lin} \sin 2
    \chi_{\rm src} \\ 0 \\
\end{array} 
\; \right] \; .
\end{eqnarray} 
\noindent We will consider both astronomical sources, which normally have
$p_{\rm src,lin} \ll 1$, and special-purpose test signals, which normally
have $p_{\rm cal,lin}=1$.

\begin{figure}[!t]
  \centering
  \includegraphics[scale=0.4]{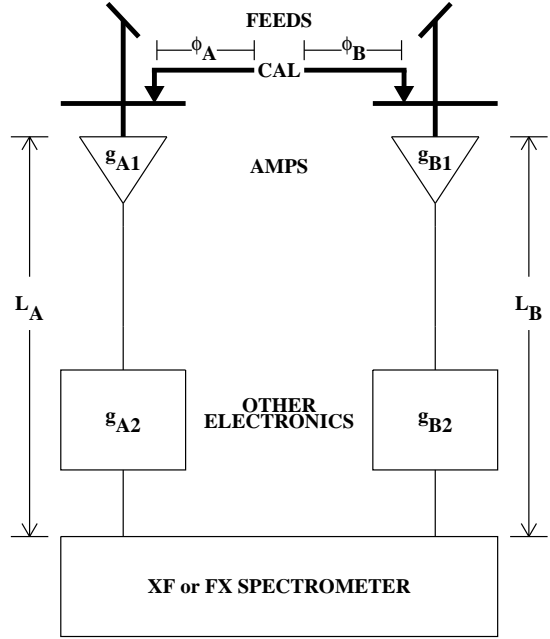}
  \vspace{-8ex}
  \caption{Block diagram of dual-polarized ($A$ and $B$) single-dish system
    (adapted from Ref.~\refcite{heiles02}). The noise diode (a.k.a.\ cal)
    output is injected through short cables and directional couplers with
    combined phase delays $\phi_A$ and $\phi_B$. The total voltage gain of
    polarization channel $A$ is $g_A=g_{A1} g_{A2}$; the voltage gains are
    complex, with an amplitude and a phase. The total cable length for
    channel $A$ is length $L_A$, which includes the run from the dish to
    the correlator input so it can be very long (more than 1 km at the
    Green Bank Telescope). The thick lines represent mechanical structures
    or passive electronics that do not change with time; the thin lines
    represent active electronics and other circuitry that do change with
    time and need calibration. \label{rh:fig:blockdiagram}}
\end{figure}

For the purpose of measuring polarization, the receiver system needs a
noise diode output that is injected into both polarizations as a correlated
calibration signal (a.k.a.~``cal''). This can be accomplished either by
injecting it externally---e.g., by a linearly polarized vertex
radiator---or by splitting the noise diode output and using two cables to
inject the signal into both polarization paths, each with a directional
coupler located just in front of its first amplifier. The position angle of
the vertex radiator should be $45$\deg\ away from the feed probes with the
ideal that the injected noise has small, or ideally zero, $Q$. The
cable-and-splitter option is the usual case, and it is depicted in the
system block diagram in \fref{rh:fig:blockdiagram}.

For the case of the cable-and-splitter injected cal, the
powers injected into the two polarization channels are almost equal, so
Stokes $Q$, which is their difference, is small and, ideally, zero;
similarly, the total polarized power fraction (\eref{rh:eq:polfrac}) is
unity. So for this ideal case, the Stokes cal vector is
\begin{eqnarray} \label{rh:eq:s_cal}
\underline{S_{\rm cal}} = \left[ 
\begin{array}{c} 
    I_{\rm cal}  \\ Q_{\rm cal}  \\ I_{\rm cal} \cos \Delta \phi_{\rm cal} \\ 
    I_{\rm cal} \sin \Delta \phi_{\rm cal} \\
\end{array} 
\; \right] \; .
\end{eqnarray} 
The angle $\Delta \phi_{\rm cal}= \phi_{{\rm cal}, A} - \phi_{{\rm cal},
  B}$ represents the phase difference between the injected noise diode
signals. A primary contributor to this difference is the different
lengths of the two noise diode cables, which makes $\Delta \phi_{\rm
  cal}$ a linear function of frequency. If the cable length difference
is exactly zero, and if the directional couplers and other devices in
the circuit are perfectly matched for the two polarizations, then this
injected cal signal is 100\% polarized with $\Delta \phi_{\rm cal}=0$
and Stokes $U_{\rm cal} = I_{\rm cal}$.

Consider the two quantities $Q_{\rm cal}$ and $\Delta \phi_{\rm
  cal}$. The cal is injected through a power splitter, cables, and
directional couplers. These are all mechanical devices and should be
stable over long periods of time; indeed, we have observationally found
that to be the case. This is fortunate, because we rely on the cal as
the secondary standard for system calibration. Therefore, it is
essential to determine $Q_{\rm cal}$ and $\Delta \phi_{\rm cal}$, and
the process of determining them we call \emph{Mueller matrix
  calibration}. We discuss this process below in
\sref{rh:sec:measurement_mmc}.


\subsection{Specifying the Stokes Vector Transfer Functions by Mueller
  Matrices}
\label{rh:sec:mueller}

\Fref{rh:fig:blockdiagram} shows a block diagram of a typical
dual-polarized radioastronomical receiver. The signal from the source first
encounters the feed. The feed rotates with respect to the source: 
for an alt/az-mounted telescope observing an astronomical source, it
rotates by the parallactic angle, while for an equatorially mounted telescope
it doesn't rotate at all. If it's an injected test signal, e.g. from a
vertex radiator, one intentionally rotates the feed for
calibration purposes.\footnote{You might
  think that rotating the vertex radiator is equivalent to rotating the
  feed. That is not the case!  When you rotate the radiator, the
  transmitted signal changes, and along with it the reflections from
  portions of the telescope, such as feed legs, change. However, when you
  instead rotate the feed, the reflections of the transmitted signal remain
  unchanged.}
The rotated feed
converts the incoming electromagnetic radiation to voltages. Finally, these
voltages are amplified to levels appropriate for the input to a digital
spectrometer.

Each of these processes modifies the Stokes parameters. We can regard
each process as having a transfer function for the four Stokes
parameters. This transfer function is a $4 \times 4$ matrix, known as
the Mueller matrix. We need the Mueller matrices for the above three
processes, discussed here in the order in which the source radiation
encounters them: \begin{enumerate}

\item \textbf{FEED ROTATION.} For the rotation of the feed by angle $\chi$
  with respect to the source, the Mueller matrix is
\begin{eqnarray} \label{rh:eq:m_rho}
\underline{\underline{M_{\chi}}} = \left[ 
\begin{array}{cccc} 
 1 &     0     &     0    & 0 \\
 0 & \phantom{-}\cos 2\chi & \sin 2\chi & 0 \\
 0 & -\sin 2\chi & \cos 2\chi & 0 \\
 0 &     0     &    0     & 1 \\
\end{array} 
\; \right] \; .
\end{eqnarray} 
The central $2 \times 2$ submatrix is, of course, nothing but a rotation
matrix.\footnote{A reminder that we adopt the notation $\underline{A}$ to
  represent a $1\times4$ column matrix and $\underline{\underline{A}}$ to
  represent a square $4\times4$ matrix.}  When the telescope rotates with
respect to the source, which is the operation described by
\eref{rh:eq:m_rho}, it is equivalent to keeping the feed stationary and
having a purely linearly polarized source emitting with position angle
$(\chi_{\rm src}-\chi)$:
\begin{eqnarray} \label{rh:eq:s_src_rho}
\underline{S_{\rm src,\chi}} = \underline{\underline{M_\chi}} 
  \cdot \underline{S_{\rm src}} = 
I_{\rm src} \cdot \left[ 
\begin{array}{c} 
    1 \\ 
    p_{\rm src,lin} \cos (2 [\chi_{\rm src} - \chi]) \\ 
    p_{\rm src,lin} \sin (2 [\chi_{\rm src} - \chi]) \\ 0 \\
\end{array} 
\; \right] \; .
\end{eqnarray} 
Note that we can regard both $\underline{S_{\rm src}}$ and
$\underline{S_{\rm src,\chi}}$ as true Stokes vectors in the sense of
\eref{rh:eq:s_def} as long as we specify that each has its own
reference coordinate system. In terms of the source's reference system,
$\underline{S_{\rm src,\chi}}$ is a pseudo-Stokes vector because its
elements are not $[I,Q,U,V]$.

\vspace{1ex}
\item \textbf{FEED COUPLING.} Next comes the feed.  Here we consider
  perfect dual-polarized feeds with native-linear or native-circular
  polarization, where the term `perfect' means that the two
  polarizations are orthogonal, the two polarizations are either purely
  linear or purely circular, and there are no losses.  We consider these
  extremes for several reasons: (1) many feeds are, in fact, close to
  perfection; (2) the discussion can focus on fundamentals without the
  fog of excess detail; (3) in practice, when you're sitting at the
  telescope and want to know how well things are working, a quick and
  approximate assessment of the receiver system is often adequate.

  \hspace{0.2in} The feed's Mueller matrix must be obtained from its Jones
  matrix.  The Jones and Mueller matrices for the general case of imperfect
  feeds are given by Eqs.~(10)--(11) of
  Ref.~\refcite{heilespnlbgloetal01}. For perfect feeds of arbitrary
  polarization, the matrices depend on two angles, called
  $\alpha_{\rm feed}$ and $\chi_{\rm
    feed}$. Ref.~\refcite{heilespnlbgloetal01} uses
  $\tan \alpha_{\rm feed}$ to specify the voltage coupling between the
  input $E$-field and output voltages and $\chi_{\rm feed}$ to represent
  the phase of that coupling (not to be confused with the position angle
  $\chi$ used in the current chapter). Perfect native linear feeds have
  $\alpha_{\rm feed} = 0\deg$ and perfect native circular feeds have
  $\alpha_{\rm feed} = \pm 45\deg$ and $\chi_{\rm feed} = \pm 90\deg$.

  Our two feed types are: 
  \begin{enumerate}

  \item \textbf{Native-Linear Feeds.} The Mueller matrix for a perfect
    native-linear feed whose probes are aligned with the azimuth and
    elevation directions is just the unitary matrix, i.e.,
    \begin{equation} \label{rh:eq:m_lin_0}
    \underline{\underline{M_{F,{\rm lin}}}} = \underline{\underline{I}} \,.
    \end{equation}
    More generally, if the native-linear feed is mounted at angle $\chi_F$
    with respect to being aligned, the Mueller matrix is simply
    \begin{eqnarray} \label{rh:eq:m_lin_rho}
    \underline{\underline{M_{F, \chi}}} = \left[
    \begin{array}{cccc} 
    1 &     0     &     0    & 0 \\
    0 & \phantom{-}\cos 2\chi_F & \sin 2\chi_F & 0 \\
    0 & -\sin 2\chi_F & \cos 2\chi_F & 0 \\
    0 &     0     &    0     & 1 \\
    \end{array} 
    \; \right] \; .
    \end{eqnarray} 

  \item \textbf{Native-Circular Feeds.} The Mueller matrix for a perfect
    native-circular feed is
    \begin{eqnarray} \label{rh:eq:m_circ}
    \underline{\underline{M_{F,{\rm cir}}}} = \left[ 
    \begin{array}{cccc} 
    \ 1 & \phantom{\mp}0 &\ 0 & \phantom{\pm}0 \\
    \ 0 & \phantom{\mp}0 &\ 0 & \pm 1 \\
    \ 0 & \phantom{\mp} 0 &\ 1 & \phantom{\pm}0 \\
    \ 0 & \mp1 &\ 0 & \phantom{\pm}0 \\
    \end{array} 
    \; \right] \; ,
    \end{eqnarray} 
    \end{enumerate}
where the signs depend on the values of $\alpha_{\rm feed}$ and $\chi_{\rm
  feed}$; the case $\alpha_{\rm feed}=+45\deg$ and $\chi_{\rm feed} =
+90\deg$ has the signs on top (i.e., $+1$ in the second row and $-$1 in the
fourth row).

\vspace{1ex}
\item \textbf{AMPLIFICATION AND ELECTRONICS.} The Mueller matrix for the
  electronics chains deals with amplitude, so we must define our
  intensity units.  First, our uppercase $G$ means power gain (which has
  no phase), while the lowercase $g$ means voltage gain, which is
  complex; $G = g\,\overline{g}$.  Following
  Ref.~\refcite{heilespnlbgloetal01}, we assume that good, but not
  perfect, intensity calibration has been previously applied to the two
  polarization channels so that the Stokes parameters have the correct
  units (e.g., temperature), and, in addition, that the total
  intensity, Stokes $I$, has perfect intensity calibration (to simplify
  the following equations). Then we define $(G_A, G_B)$ to be the power
  gains for the two polarization channels. Because of our assumptions we
  write $G_A = (1 + \delta G)$ and $G_B = (1 - \delta G)$, where $\delta
  G$ is unitless and $|\delta G| \ll 1$. For consistency with
  Ref.~\refcite{heilespnlbgloetal01}, we define $\Delta G_{AB} = 2
  \delta G$. Then, to first order in $\Delta G_{AB}$, the Mueller matrix
  for the
  electronics chains (see \fref{rh:fig:blockdiagram}) is
    \begin{eqnarray} \label{rh:eq:m_amp}
    \underline{\underline{M_{AB}}} = \left[ 
    \begin{array}{cccc} 
    1                &  \frac{\Delta G_{AB}}{2}   &     0    &      0          \\
    \frac{\Delta G_{AB}}{2}  &          1           &     0    &      0          \\
    0                &       0              & \cos \Delta \phi_{AB} & -\sin \Delta \phi_{AB}  \\
    0                &       0              & \sin \Delta \phi_{AB} & \phantom{-}\cos \Delta \phi_{AB}  \\
    \end{array} 
    \; \right] \; .
    \end{eqnarray} 
The two parameters in
$\underline{\underline{M_{AB}}}$ are the relative power gain ($\Delta
G_{AB}$) and phase delay between the two polarization channels ($\Delta
\phi_{AB}$) and are both associated with the electronics and the
circuitry, including cable lengths. These quantities can change with
time because they are associated with active electronics, so they need
to be measured often enough to keep up with the variability of system
electronics---and at least once per observing session.

  \end{enumerate}


\subsection{The Measured Pseudo-Stokes Vector $\underline{{\cal S}^{\rm cor}}$ 
for Several Cases}

After being operated on by these three Mueller matrices, the original
source Stokes vector becomes the previously defined pseudo-Stokes
vector, producing voltages $V_A$ and $V_B$ at the input to the
correlator. The correlator generates the auto- and cross-products as
discussed above, producing the pseudo-Stokes vector output
$\underline{{\cal S}^{\rm cor}}$.

When the system looks at `blank sky', the input noise is mainly from the
receiver, with a small contribution from the sky and ground pickup. For
purposes of illustration, we include only the receiver contribution. In
this case, the noise is injected after the feed so the only Mueller matrix
that operates is $\underline{\underline{M_{AB}}}$. Denote the associated
pseudo-Stokes vector by $\underline{{\cal S}^{\rm cor}_{\rm rx}}$:
\begin{equation} \label{rh:eq:cor_rx}
\underline{{\cal S}^{\rm cor}_{\rm rx}} = \underline{\underline{M_{AB}}} \cdot
\underline{S_{\rm rx}} \,.
\end{equation}
\noindent When on the source, with the cal off, we see
\begin{equation} \label{rh:eq:cor_srcon}
\underline{{\cal S}^{\rm cor}_{\rm src}} = \underline{\underline{M_{AB}}} \cdot
\underline{\underline{M_F}} \cdot \underline{\underline{M_\chi}} \cdot
\underline{S_{\rm src} } + \underline{{\cal S}^{\rm cor}_{\rm rx}} \,.
\end{equation}
\noindent And when off the source with the cal on:
\begin{equation} \label{rh:eq:cor_calon}
\underline{{\cal S}^{\rm cor}_{\rm cal}} = \underline{\underline{M_{AB}}}
\cdot \underline{S_{\rm cal} } + \underline{{\cal S}^{\rm cor}_{\rm rx}} \,.
\end{equation}
For an accurate measurement of the source or cal deflection, we must
subtract the off-source contribution $\underline{{\cal S}^{\rm cor}_{\rm
    rx}}$, as is usual for all single-dish measurements. Denote these
deflections with the prefix $\Delta$. Then for any type of feed, the
cable-injected cal response is
\begin{eqnarray} \label{rh:eq:delcor_calon}
\underline{\Delta{\cal S}^{\rm cor}_{\rm cal}} = 
\underline{\underline{M_{AB}}} \cdot \underline{S_{\rm cal} } = 
I_{\rm cal} \cdot \left[ 
\begin{array}{c} 
    1 \\ 
    \frac{\Delta G_{AB}}{2} + \frac{Q_{\rm cal}}{I_{\rm cal}} \\
    \cos( \Delta \phi_{AB} + \Delta \phi_{\rm cal}) \\
    \sin( \Delta \phi_{AB} + \Delta \phi_{\rm cal}) \\
\end{array} 
\; \right] \; .
\end{eqnarray} 
\noindent We have assumed $\frac{Q_{\rm cal}}{I_{\rm cal}} \ll 1$ and
  kept only first-order terms.

Similarly, for the source deflection, we get
\begin{eqnarray} \label{rh:eq:delcor_srcon}
\underline{\Delta{\cal S}^{\rm cor}_{\rm src}} =
\underline{\underline{M_{AB}}} \cdot 
\underline{\underline{M_F}} \cdot 
\underline{\underline{M_\chi}} \cdot \underline{S_{\rm src}} \; .
\end{eqnarray} 
\noindent For a perfect native-linear feed with probes aligned with the
azimuth and elevation directions,
$\underline{\underline{M_F}} = \underline{\underline{M_{F,{\rm lin}}}} = \underline{\underline{I}}$
(\eref{rh:eq:m_lin_0}) and
\begin{eqnarray} \label{rh:eq:delcor_srcon_lin}
\underline{\Delta{\cal S}^{\rm cor}_{\rm src,lin}} = 
I_{\rm src} \cdot \left[ 
\begin{array}{c} 
    1 + \frac{\Delta G_{AB}}{2} p_{\rm src,lin} \cos (2 [\chi_{\rm src}-\chi]) \\ 
    \frac{\Delta G_{AB}}{2} + p_{\rm src,lin} \cos (2 [\chi_{\rm src}-\chi]) \\ 
    p_{\rm src,lin} \cos \Delta \phi_{AB}  \sin (2 [\chi_{\rm src}-\chi]) \\
    p_{\rm src,lin} \sin \Delta \phi_{AB}  \sin (2 [\chi_{\rm src}-\chi]) \\
\end{array} 
\; \right] \; .
\end{eqnarray} 
\noindent The $\Delta {\cal S}^{\rm cor}_{\rm src,lin,0}$ element\footnote{
  $\Delta {\cal S}^{\rm cor}_{\rm src,lin,0}$ is the zeroth element of the
  $\underline{\Delta{\cal S}^{\rm cor}_{\rm src,lin}}$ pseudo-Stokes vector;
  see \eref{rh:eq:s_def}.} is the pseudo-Stokes $I$ and is not equal to
unity. This can be awkward for Mueller matrix calibration, when one almost
always forces $\Delta {\cal S}^{\rm cor}_{\rm src,lin,0}$ to be unity to
eliminate the influence of overall system gain changes.  These can occur,
for example, from pointing errors or position-dependent telescope surface
distortions and other circumstances that reduce the overall system gain. So
one must divide the other three pseudo-Stokes parameters by $\Delta {\cal
  S}^{\rm cor}_{\rm src,lin,0}$.  Fortunately, for the common case when an
astronomical source is used for the calibration, we almost always have
$p_{\rm src,lin} \ll 1$; this makes the contribution of the non-unity
portion of $\Delta {\cal S}^{\rm cor}_{\rm src,lin,0}$ to the
other three pseudo-Stokes parameters second-order, so it can be
neglected. However, for a locally generated test signal, $p_{\rm cal,lin}$
is likely to be unity. The easiest way to deal with this is to rescale the
amplitudes so that $\Delta G_{AB}$ itself becomes second order.

For a perfect native-circular feed, $\underline{\underline{M_F}}$ is given
by \eref{rh:eq:m_circ}, and
\begin{eqnarray} \label{rh:eq:delcor_srcon_cir}
\underline{\Delta {\cal S}^{\rm cor}_{\rm src,cir}} = 
I_{\rm src} \cdot \left[ 
\begin{array}{c} 
    1 \\ 
    \frac{\Delta G_{AB}}{2} \\
    \phantom{-}p_{\rm src,lin} \sin( \Delta \phi_{AB} + 2 [\chi_{\rm src}-\chi]) \\
    -p_{\rm src,lin} \cos( \Delta \phi_{AB} + 2 [\chi_{\rm src}-\chi]) \\
\end{array} 
\; \right] \; .
\end{eqnarray} 


\subsection{Discussion: The Process of `Mueller Matrix Calibration'
\label{rh:sec:measurement_mmc}}

Suppose you make a single measurement $\underline{\Delta {\cal S}^{\rm
    cor}_{\rm src}}$ of the deflection of a linearly polarized source
and wish to derive the source's linear polarization fraction $p_{\rm
  src,lin}$ and position angle $\chi_{\rm src}$ from the measured
$\underline{\Delta {\cal S}^{\rm cor}_{\rm src}}$. If all of the
off-diagonal terms in the three Mueller matrices were zero, this would
be easy. However, this is never the case. If you know the three Mueller
matrices, then you can calculate the inverse of their matrix product and
derive $\underline{S_{\rm src}}$ from the measured $\underline{\Delta
  {\cal S}^{\rm cor}_{\rm src}}$ using \eref{rh:eq:delcor_srcon};
alternatively, if you know $p_{\rm src,lin}$ and $\chi_{\rm src}$
(because it's a polarization calibration source, for example), then you
can analytically calculate $\underline{\Delta {\cal S}^{\rm cor}_{\rm
    src}}$ from $\underline{S_{\rm src}}$ using
\eref{rh:eq:delcor_srcon}. Either way, the parameters $\Delta
G_{AB}$ and $\Delta \phi_{AB}$ need to be known.  To determine them
we need to use the calibration noise diode, which produces the
deflection given by \eref{rh:eq:delcor_calon}. This deflection
depends on four quantities: our two required amplifier-chain parameters
$\Delta G_{AB}$ and $\Delta \phi_{AB}$ (which change with time), and
the two cal-injection parameters $Q_{\rm cal}$ and $\Delta \phi_ {\rm
  cal}$ (which do not change with time).

We cannot determine $\Delta G_{AB}$ and $\Delta \phi_{AB}$ without
knowing $Q_{\rm cal}$ and $\Delta \phi_{\rm cal}$. We call the
process of determining $Q_{\rm cal}$ and $\Delta \phi_{\rm cal}$ the
\emph{Mueller matrix calibration}. Mueller matrix calibration is done by
observing a polarization calibrator with known intensity and
polarization to obtain $\underline{\Delta {\cal S}^{\rm cor}_{\rm src}}$
over a range of parallactic angle $\chi$ and, in addition, obtaining the
cal deflection $\underline{\Delta {\cal S}^{\rm cor}_{\rm cal}}$.  One
then plots the $\chi$-dependence of the four elements of
$\underline{\Delta {\cal S}^{\rm cor}_{\rm src}}$. The first element,
Stokes $I$, is constant by definition, because we always deal with
fractional Stokes parameters.  The remaining three elements vary
periodically with $\chi$, and from the amplitudes and phases of their
variation one can use least-squares fitting of
\eref{rh:eq:delcor_srcon_lin} or \eref{rh:eq:delcor_srcon_cir} to
derive all of the parameters.

Least-squares fitting is best for accuracy, but referring to that process
does not aid our phenomenological understanding. We can develop our
understanding by solving for the parameters using basic algebra.  First,
obtain 
$\Delta \phi_{AB}$ and $\Delta \phi_{\rm cal}$ from
\eref{rh:eq:delcor_srcon_lin} and \eref{rh:eq:delcor_calon}:\footnote{
  $\Delta {\cal S}^{\rm cor}_{{\rm src,lin,}i}$ is the $i$th element of the
  $\underline{\Delta{\cal S}^{\rm cor}_{\rm src,lin}}$ pseudo-Stokes vector;
  see \eref{rh:eq:s_def}.}
\begin{equation}
  \begin{split}
  \Delta \phi_{AB} &= \tan^{-1}  \left( \frac{\Delta {\cal S}^{\rm cor}_{\rm src,lin,3}}
{\Delta {\cal S}^{\rm cor}_{\rm src,lin,2}} \right) \,,\\
  \Delta  \phi_{AB} + \Delta \phi_{\rm cal} &= \tan^{-1} 
\left( \frac{\Delta {\cal S}^{\rm cor}_{\rm cal,3}}
{\Delta {\cal S}^{\rm cor}_{\rm cal,2}} \right) \,.
\end{split}
\end{equation}
\noindent Next, plot $\Delta {\cal S}^{\rm cor}_{\rm src,lin,1}$ versus
$\chi$. The part that varies with $\chi$ gives $p_{\rm src,lin}$ and the
offset of this cosine wave from zero gives $\Delta G_{AB}$. Combine this
with $\Delta {\cal S}^{\rm cor}_{\rm cal,1}$ to obtain $Q_{\rm cal}$.  One
assumes, of course, that during the time interval for this calibration the
parameters stay fixed---in particular, that the electronics parameters
$\Delta G_{AB}$ and $\Delta \phi_{AB}$ stay fixed. Experience shows that
with modern electronics at the 305-m Arecibo telescope and the 100-m Green
Bank Telescope (GBT) this assumption is good.

\begin{figure}[!t]
  \centering
  \includegraphics[width=2.45in] {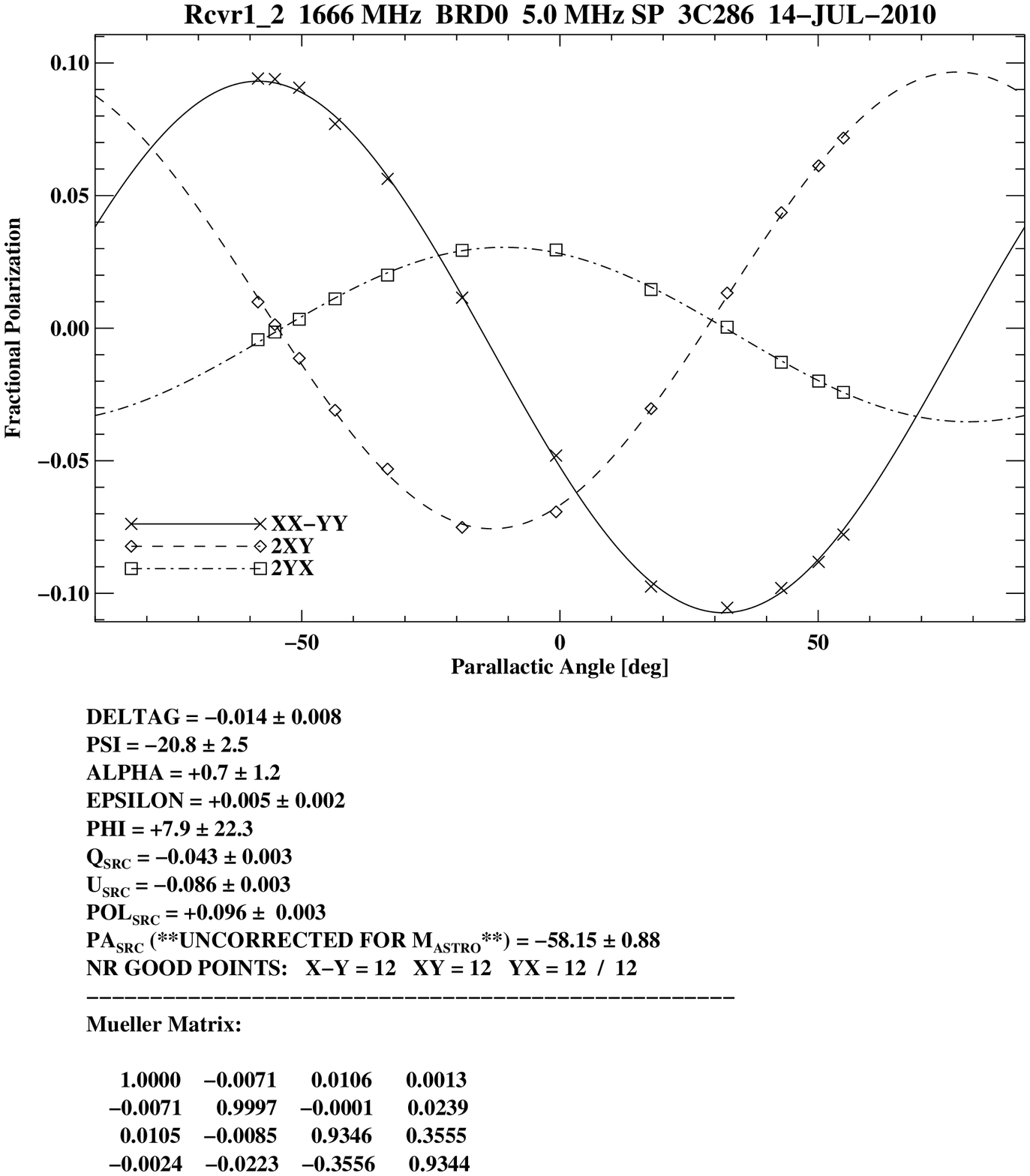}
  \includegraphics[width=2.45in] {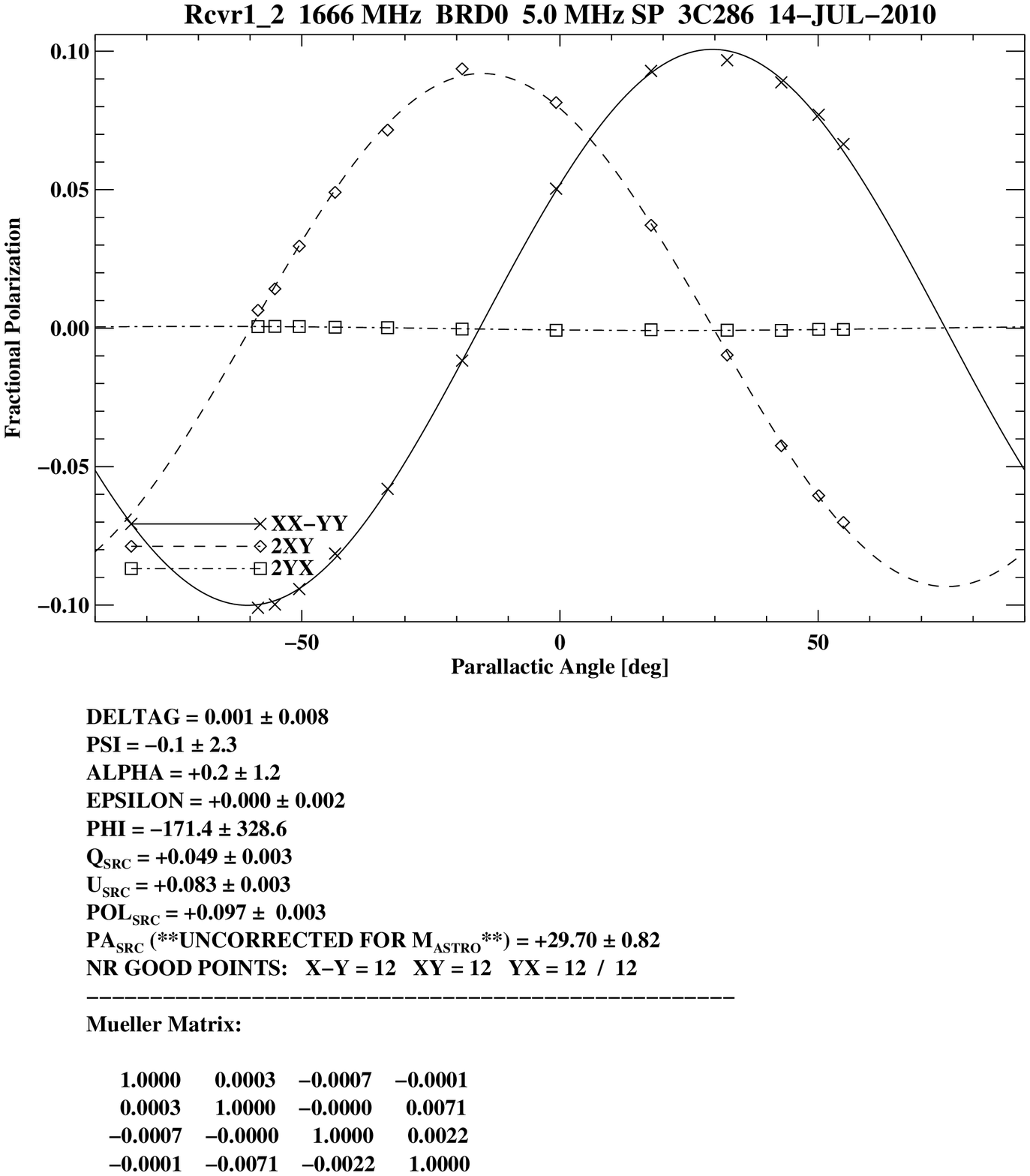}
  \vspace{-8ex}
  \caption{(\textit{Left}) Mueller matrix calibration of the native-linear
    $L$-band GBT receiver showing the normalized $\underline{{\cal S}^{\rm cor}_{\rm
      src,lin}}$ outputs versus parallactic angle $\chi$.  The crosses
    (solid line) show $({ \Delta {\cal S}^{\rm cor}_{\rm src,
        lin,1}})/({\Delta {\cal S}^{\rm cor}_{\rm src, lin,0}})$, the diamonds
    (dashed line) show $({\Delta {\cal S}^{\rm cor}_{\rm src,
        lin,2}})/({\Delta {\cal S}^{\rm cor}_{\rm src, lin,0}})$, and the
    squares (dash-dot line) show $({ \Delta {\cal S}^{\rm cor}_{\rm src,
        lin,3}})/({\Delta {\cal S}^{\rm cor}_{\rm src, lin,0}})$.  Results
    of the least-squares fit are given below the plot (see text).
    (\textit{Right}) The same plot after the 3C~286 data have been
    corrected by the derived Mueller matrix. The same least-squares fit
    process was performed on the calibrated data; the leakage of Stokes
    parameters has been minimized, as can be seen from the plots and from
    the near-zero off-axis terms in the Mueller matrix derived from these
    Mueller-matrix-corrected data.
 \label{rh:fig:mueller}}
\end{figure}

\Fref{rh:fig:mueller} shows a set of 1666 MHz Mueller matrix
calibration data from the famous polarization calibrator 3C~286 for the
native-linear polarization system at the GBT. The crosses (solid line) show
$(\Delta {\cal S}^{\rm cor}_{\rm src, lin,1})/(\Delta {\cal S}^{\rm cor}_{\rm src,
    lin,0})$ and the diamonds (dashed line) show $(\Delta {\cal S}^{\rm
    cor}_{\rm src, lin,2})/({\Delta {\cal S}^{\rm cor}_{\rm src, lin,0}})$.  If
the data were perfectly calibrated for polarization, these two outputs
would equal $Q_{\rm src}$ and $U_{\rm src}$ and would vary sinusoidally
with twice the parallactic angle, with the two sinusoids having equal
amplitude and no offsets from zero.  This is definitely not the case.  The
squares (dash-dot line) in \fref{rh:fig:mueller} represent $(\Delta {\cal S}^{\rm cor}_{\rm src,
    lin,3})/(\Delta {\cal S}^{\rm cor}_{\rm src, lin,0})$ and 
 reveal a major leakage of linear polarization into
Stokes $V$.  A nonlinear least-squares fit of these data yields the first
seven parameters listed below the left plot in 
\fref{rh:fig:mueller}.\footnote{In \fref{rh:fig:mueller}, the first
  two parameters are labelled DELTAG and PSI and correspond to our
  $\frac{Q_{\rm cal}}{I_{\rm cal}}$ and $\phi_{\rm cal}$; the next three
  deal with feed imperfections; and the last four are the source
  polarization. For a detailed description of all the listed parameters,
  see Sec.\ 7.1 of Ref.~\refcite{heilespnlbgloetal01}.} The associated Mueller
matrix is listed at the bottom of the left panel. The nonzero off-axis
elements quantify the leakage of one uncalibrated Stokes parameter into
another. If $\underline{{\cal S}^{\rm cor}_{\rm src,lin}}$ is corrected by this
Mueller matrix, the proper $\chi$-dependencies of the elements of
$\underline{{\cal S}^{\rm cor}_{\rm src,lin}}$ are recovered, as depicted in the
right panel of \fref{rh:fig:mueller}.

\subsection{Two Important Subtleties Regarding Relative Phase $\phi_{AB}$}

\subsubsection{System Cable Lengths} 
\label{rh:sec:cables}

Various electronics components in the signal path between the feed and
the correlator introduce complex voltage gains that can include
amplification, attenuation, and phase changes (e.g., some amplifiers
introduce a phase shift of 180\deg).  Of particular importance: the
combined lengths of the coaxial cables and optical fibers differ between
the two signal paths ($L_A$ and $L_B$ in 
\fref{rh:fig:blockdiagram}).  Environmental factors can cause these
lengths to change with time.  A difference between the path lengths
produces a phase difference in radians of
\begin{equation}
\delta\phi_{AB} = \frac{2\pi(L_{A} - L_{B})}{\lambda}\,,
\end{equation}
and this phase difference depends on frequency as
\begin{equation}
\frac{d\delta\phi_{AB}}{d\nu} = \frac{2\pi(L_{A} - L_{B})}{c}\,.
\end{equation}
This phase difference, $\delta \phi_{AB}$, adds to other
contributions to produce the total phase difference $\Delta
\phi_{AB}$. Measured values of the total phase gradient
$\frac{d\Delta\phi_{AB}}{d\nu}$ at Arecibo and the GBT are about
0.3 rad MHz$^{-1}$, corresponding to a difference in cable/fiber length
of $\sim$20~m.  This is surprisingly large, even considering the extreme
distances between the feed and correlator for these telescopes.  


\subsubsection{System Band-Limiting Filters and Their Induced Kramers-Kronig Phase
  Shifts} 
\label{rh:sec:filters}

At some point in the receiver chain one always has a band-limiting
filter. Frequency-dependent gains automatically introduce phase delays,
which can be calculated from the Kramers-Kronig relations. If the
filters in the two polarizations are not perfectly matched, a
frequency-dependent phase difference between the  two polarization
channels ensues.  This can be particularly
serious when the filters have significant gain changes within the usable
portion of the band.

The exact formula for the phase shift (in radians) induced by a power
gain change in an electrical circuit is given by Eq.~(2) of
Ref.~\refcite{bode40}:\footnote{You would miss a lot if you pass up the
  opportunity to read Bode's paper\citep{bode40}, particularly the first
  six pages. Go to
  \url{http://www.alcatel-lucent.com/bstj/}\,. N.B.: Bode's derivation
  treats changes in logarithmic attenuation ${\cal A}$; since we're
  treating changes in logarithmic power gain ${\cal G}$, we've set 
${\cal G}= -{\cal A}$ in his equations.}
\begin{equation}
\phi(\nu_c) = -\frac{1}{\pi} \int_{-\infty}^{\infty} \frac{d{\cal G}(u)}{du}
\ \ln \left[ \coth \left( \frac{\left|u\right|}{2}\right) \right] du \,,
\end{equation}
where ${\cal G}(u)$ is the filter power gain in nepers, $u =
\ln \left( \frac{\nu}{\nu_c} \right)$, $\nu$ is frequency, and $\nu_c$
is an arbitrarily chosen frequency. The weighting function $\ln \left[
  \coth \left( \frac{\left|u\right|}{2}\right) \right] $ is sharply
peaked at $u=0$ where $\nu=\nu_c$, so a good approximation eliminates
the integral and uses only the local derivative (Eq.~22 of
Ref.~\refcite{odonnelljm81}):
\begin{equation} \label{rh:eq:approxphase}
\phi(\nu_c) = \left. -\frac{\pi}{2} \frac{d{\cal G}(u)}{du} \right|_{u=0}
\,.
\end{equation}
\noindent Thus a non-flat filter produces phase shifts.

\begin{figure}[!t]
\begin{center}
\includegraphics[width=0.48\textwidth]{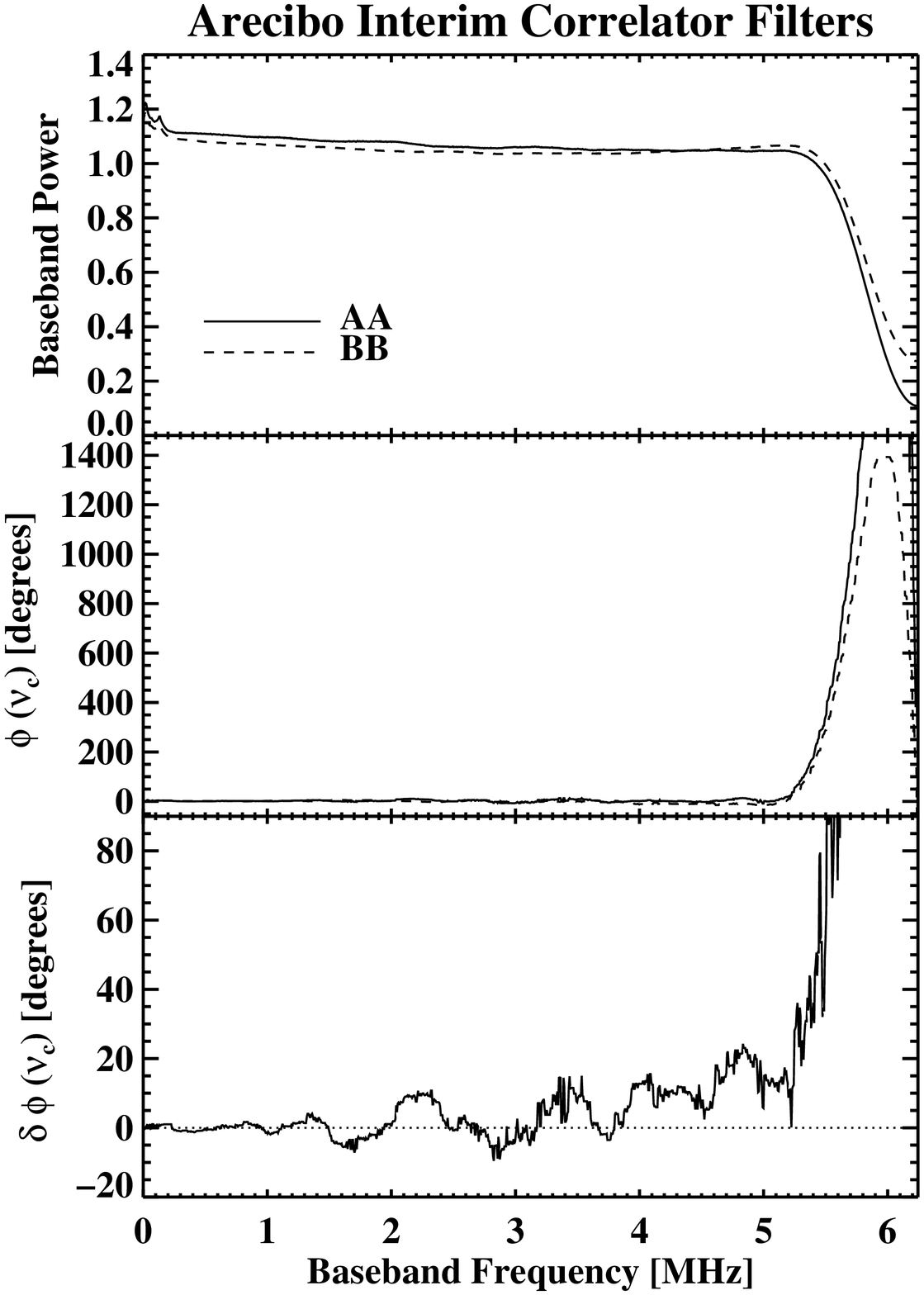}
\includegraphics[width=0.48\textwidth]{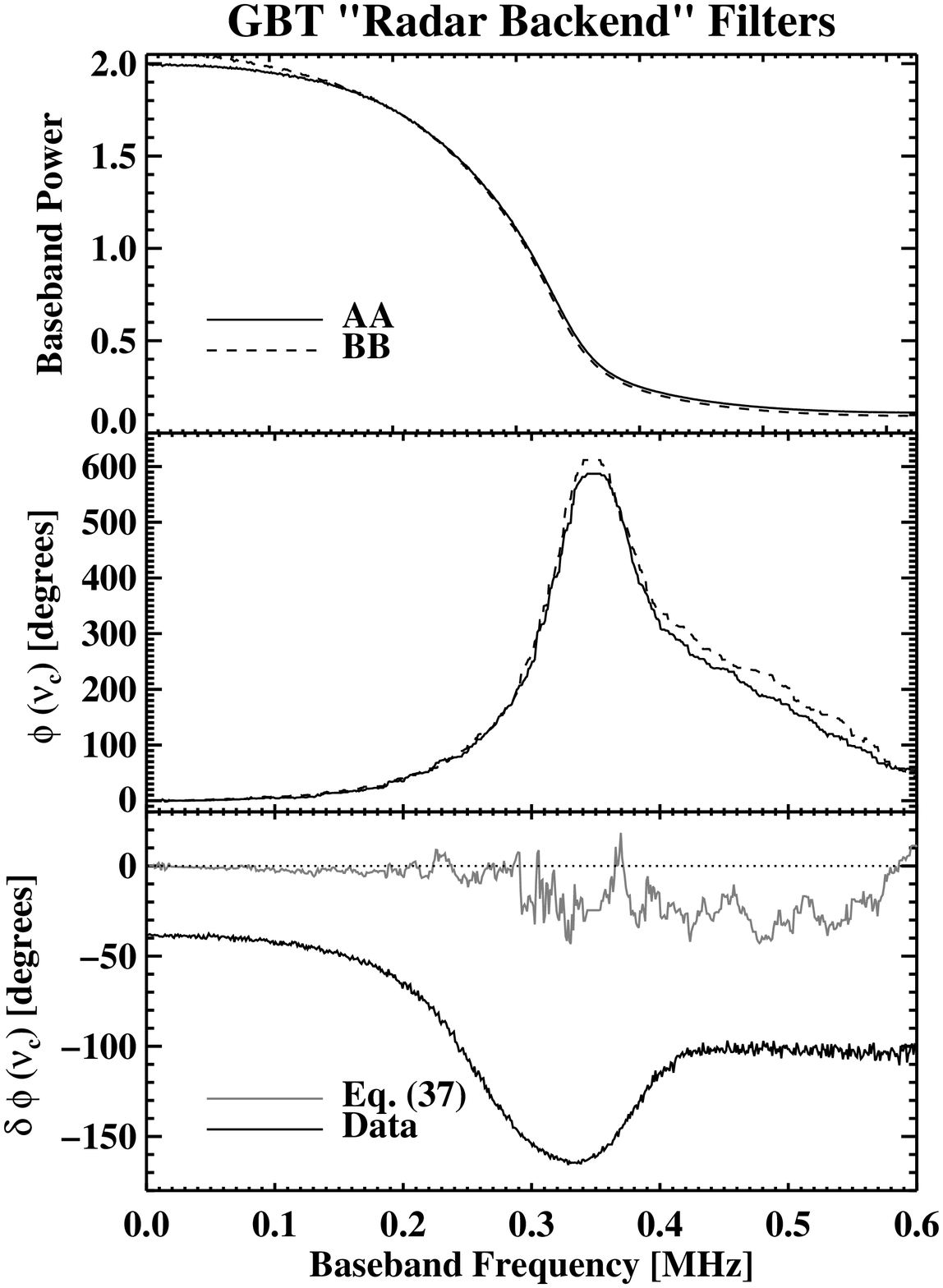}
\end{center}
\vspace{-4ex}
\caption{\footnotesize Filter shapes and their theoretical phase delays for
  the Arecibo Interim Correlator (\textit{left panel}) and the GBT ``Radar
  Backend'' (\textit{right panel}). Both are low-pass baseband filters with
  complex digital sampling, so the frequency coverage extends from $-B$ to
  $+B$, where $B$ is the cutoff frequency. In the bottom panel for the GBT,
  the smoother curve is the measured phase difference and the gray noisy
  curve is the theoretical one from \eref{rh:eq:approxphase}.
  \label{rh:fig:filters}}
\end{figure}

The left panel of \fref{rh:fig:filters} depicts the power gains and phase
delays for Arecibo's interim correlator, for which the baseband low-pass
filters (cutoff frequency 6.25 MHz) are digitally defined and are
remarkably flat. We show only the positive-frequency half. Phase shifts
occur only at the high-frequency end, where the responses of the filters
drop precipitously.

In contrast, the GBT Radar Backend filters (\fref{rh:fig:filters},
right-hand panel) fall to zero gradually, with no sharp cutoff
frequency. Thus, the power gain varies rapidly within the observing band (top
panel), with correspondingly large (huge!) phase delays (middle panel),
peaking at $\sim$600\deg! With such large phase delays, even small
differences between the $A$ and $B$ filters lead to significant
frequency-dependent relative phase delays $\delta \phi_{AB}$ (bottom
panel). For native-linear polarization, these delays interchange power
between $U$ and $V$; for native-circular, they interchange $Q$ and $U$.
These phase differences {\it must be corrected}.

For the GBT, the bottom plot shows both the theoretical (the noisy curve, from
\eref{rh:eq:approxphase}) and measured (the smoother curve, from
correlated noise injection) phase differences between the two signal
paths. The theory and the data do not agree at
all. The reason is inaccuracy in the filter shape resulting from
uncorrected 4-bit quantized voltage sampling. Specifically, the
calculated filter responses do not fall to zero at high frequencies, as
they actually do. If, as a numerical experiment, we displace the $BB$
curve downwards by 0.03, the theory curve becomes equal to the
smoother measured one above 0.35 MHz. Thus, the theory curve is
inaccurate and noisy, because it is the difference between two large
numbers, neither of which is itself very accurate.



\section{Off-Axis Instrumental Polarization}
\label{rh:sec:offaxisinstrumental}

Thus far we've considered the polarization properties of radiation entering
the feed along the optical axis of a telescope's main beam.  However,
radio telescopes pick up radiation off-axis via sidelobe response.  The
polarization state of incoming radiation can be altered in these polarized
sidelobes (and even inside the main beam!) in such a way that unpolarized
astronomical radiation can be converted to a polarized response affecting
the on-axis signal.

Understanding the mechanisms that create this off-axis instrumental
polarization is the domain of antenna engineers whose interests lie in
building efficient dual-polarized communication systems that carry a pure
polarized signal (what they call the \textit{co-polarized} signal) in one
channel without allowing that information to leak into the orthogonal
polarization state (what they call a \textit{cross-polarized}
signal).\footnote{Ref.~\refcite{carozziw11} lists the various terms that
  engineers and astronomers use for the singular concept of instrumental
  polarization, among them: cross-polarization, feed or polarization
  leakage, $D$-terms, cross-coupling, mutual coupling, cross-talk, barrel
  distortion, and beam squash.}  In order to accomplish this, the two
$E$-field polarization states (horizontal and vertical for a dual-linear
feed, RCP and LCP for a dual-circular feed) need to be perfectly
orthogononal across the aperture plane of the telescope. This is an
impossible task: there is always some cross-polarization inherent in the
system.  We investigate below some of the most common causes of this
cross-polarization from both the engineer's viewpoint of transmitting from
the focus and the astronomer's reciprocal perspective of receiving at the
focus.


\subsection{Cross-Polarization Induced by the Feed and Dish Surface: Beam Squash}
\label{rh:sec:instrumental.pol.squash}

Ref.~\refcite{collin85} uses multiple methods to analytically derive the
cross-polarization response of a circularly symmetric paraboloidal
reflector with a feed located at the primary focus.  The resulting
cross-polar pattern depends on the analysis method, but two components are
always present: a depolarization pattern caused by the curvature of the
reflector surface and a pattern from the inherent cross-polarization of the
feed.  Both contributions will produce $E$-field aperture distributions
with nulls along the principal planes\footnote{For a dual-linear feed, the
  two principal planes are those that contain the reflector axis and the
  orthogonal feed probes.} and field maxima in the $\pm45\deg$
planes.\citep{stutzmant13,teradas96,tinbergen96,collin85} The far-field
$E$-field radiation pattern can then be produced from this $E$-field
aperture distribution via 2-D Fourier transform
integration.\citep{stutzmant13}

Astronomers are interested in knowing how their telescope responds to an
unpolarized source of radiation at any angle off of the optical axis.  This
can be measured in practice for a single-dish telescope by mapping out the
Stokes parameter response of a strong unpolarized continuum source as the
main beam is driven around an area centered on the source.  Fractional
Stokes parameter beam maps are then generated by dividing these Stokes beam
maps by $I_{\rm peak}$, the peak Stokes $I$ response of the main
beam.\footnote{If one wanted to estimate the instrumental contribution to
  the on-axis Stokes $Q$ response from an unpolarized source in the first
  sidelobe, one would multiply the source's Stokes $I$ brightness by the
  fractional polarization at the appropriate location in the
  $Q/I_{\rm peak}$ pattern.  These fractional Stokes parameter beam maps
  should not be confused with point-for-point maps of fractional Stokes
  parameters, e.g., the $Q$ pattern divided by the $I$ pattern.  While an
  interferometer might be able to measure such a pattern readily, a
  single-dish telescope does not have enough dynamic range or angular
  resolution to quickly measure point-for-point fractional polarization in
  far-out sidelobes.}  For notational efficiency, we will refer to the
fractional quantities $\{I,Q,U,V\}/I_{\rm peak}$ in the remainder of this
section as simply $\{I,Q,U,V\}$.

\begin{figure}[!tbp]
  \centering
  \includegraphics[width=1.0\textwidth]{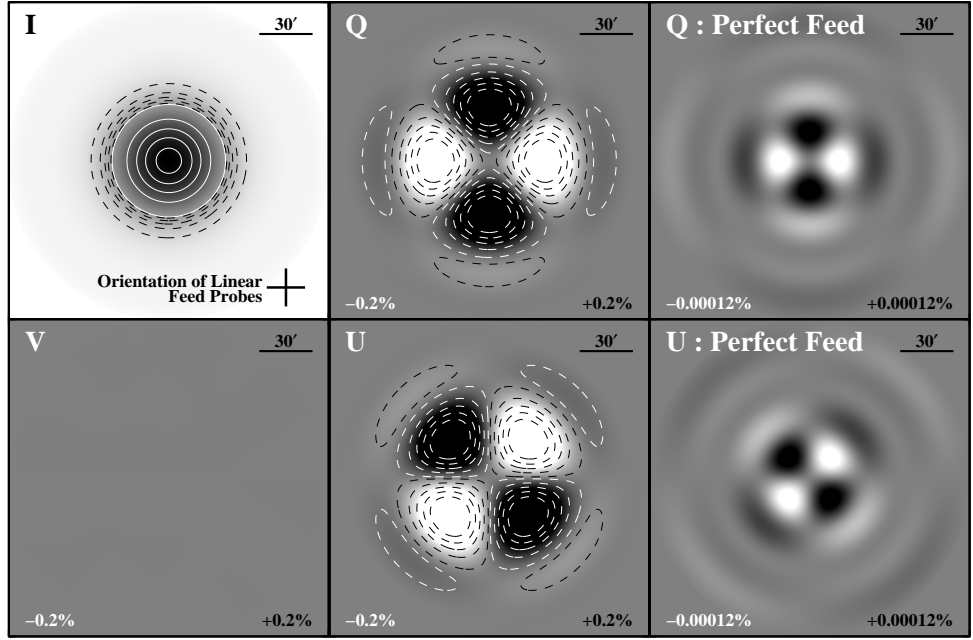}
  \caption{GRASP-generated far-field beam patterns at 1420 MHz for a
    circularly symmetric paraboloidal reflector of diameter 25.9 m and
    $f/D=0.2941$ with no feed legs or aperture blockage. All beam patterns
    are normalized to the peak main-beam Stokes $I$ response and each frame
    covers 3\deg$\times$3\deg\ on beam center.  A simulated feed pattern
    for a circular-waveguide feed (with inherent cross-polarization) with
    four $\lambda$/4 chokes and dual-linear probes was used to illuminate
    the primary, producing beam patterns for: \textit{(top left)} Stokes
    $I$ with grayscale covering 0--100\% (white to black), solid white
    contours covering (10\%, 30\%, 50\%, 70\%, 90\%), and dashed black
    contours covering (1\%, 3\%, 5\%, 7\%, 9\%); Stokes $Q$ \textit{(top
      middle)}, $U$ \textit{(bottom middle)}, and $V$ \textit{(bottom
      left)} with grayscale covering (white to black) $\pm0.2$\% of the
    peak Stokes $I$ (thus white is $-0.2$\%, black $+0.2$\%, and gray 0\%),
    dashed black contours covering ($-$0.02\%, $-$0.10\%, $-$0.18\%,
    $-$0.26\%, $-$0.34\%, $-$4.2\%), dashed white contours covering
    (0.02\%, 0.10\%, 0.18\%, 0.26\%, 0.34\%, 4.2\%), and 0\% contour
    omitted.  The orientation of the dual-linear feed probes is indicated;
    the lobes of the Stokes $Q$ pattern align with the probes while the $U$
    pattern is oriented at $45\deg$.  There is no discernible $V$ response.
    The Stokes $Q$ \textit{(top right)} and $U$ \textit{(bottom right)}
    patterns are also shown for the same primary reflector being fed by a
    perfect feed with no inherent cross-polarization.  Grayscale covers
    $\pm0.00012$\% (white to black).  These patterns show that the
    cross-polarization induced by the reflector alone has the same
    character and orientation as that produced by the waveguide feed and
    reflector working in conjunction, but the reflector-only pattern is
    narrower and more than 1000 times weaker.}
  \label{rh:fig:ng.beam.maps}
\end{figure}

Before inspecting a measured polarized beam pattern, we can investigate
what one might expect from a perfect telescope.  For the last few decades,
the commercial software package GRASP has developed into a sophisticated
tool allowing the far-field vector $E$-field response of reflector antennas
to be precisely modelled using efficient algorithms for physical optics and
the physical theory of diffraction.  We follow the lead of
Ref.~\refcite{nglcrgrv05} and use GRASP to model the transmitted far-field
pattern of the circularly symmetric DRAO 25.9-m diameter paraboloidal
telescope ($f/D=0.2941$) fed from the primary focus by a simulated feed
pattern for a circular-waveguide feed (with inherent cross-polarization)
with four $\lambda$/4 chokes and dual-linear probes.\citep{wohllebenml72}
The Stokes parameters were constructed from the simulated far-field
$E$-field distribution for a given orientation of the feed probes via
\eref{rh:eq:stokes.orthogonal.cartesian}. (To simplify the modelling even
further, we exclude any feed-support legs and aperture blockage.)  Then,
invoking the principle of reciprocity, the feed was rotated through
180\deg\ and each of the Stokes patterns averaged over these orientations
to simulate the transmission of unpolarized light in the far field.
\Fref{rh:fig:ng.beam.maps} shows these averaged fractional Stokes parameter
beam patterns, which also represent the telescope's response to unpolarized
radiation.  The rightmost panels show the simulated beam patterns for a
perfect linear dipole feed transmitting onto the same reflector geometry;
this feed has absolutely no inherent cross-polarization, so that any Stokes
$Q$ or $U$ response will be entirely brought about by the reflector
surface.  Some important properties are immediately evident:
\begin{enumerate}
 
\item The Stokes $Q$ and $U$ cross-polarization patterns resemble a
  four-lobed clover leaf with lobes on opposite sides of the beam center
  having identical signs; the signs of adjacent lobes alternate in beam
  azimuth.  We call this pattern \textit{beam squash}. For a Stokes $Q$
  pattern with its positive-response lobes aligned along the vertical axis,
  this is equivalent to the beamwidth being larger in the vertical
  direction than in the horizontal direction (meaning the feed pattern
  illuminating the primary reflector is wider in the horizontal direction
  than in the vertical).

\item The lobes of the beam squash pattern are aligned with the feed probe
  orientation for Stokes $Q$ and are aligned at $45\deg$ for Stokes $U$.

\item The sign of the beam squash response reverses between the main beam
  and the first sidelobe.

\item The beam squash produced by the dish is dwarfed (by a factor of 3000
  in this instance) by the beam squash inherent in the feed
  response.\footnote{The reflector cross-polarization decreases with
    increasing $f/D$.\citep{stutzmant13,ludwig73}} This situation almost
  always obtains\citep{collin85,baars07}, even for corrugated conical horns
  whose cross-polarization response can be designed to be significantly
  smaller than other types of feed.\citep{clarricoatso84,collin85}

\end{enumerate}

\subsection{Polarization Induced by the Feed Location: Beam Squint}
\label{rh:sec:instrumental.pol.squint}

If a feed is tilted or displaced from the focus of a reflector such that
the feed axis and the reflector axis are misaligned, an amplitude or phase
slope is induced across the reflector's aperture plane.  In the far-field
response, this translates to the RCP and LCP beams pointing in slightly
different directions on either side of boresight; the displacement occurs
in the plane that is orthogonal to the plane of symmetry of the reflector
and is known as \textit{beam
  squint}.\citep{chut73,rudgea78,duanr91,teradas96} So if a feed is tilted
and/or displaced from the reflector axis in the azimuth direction, the beam
squint lobes will lie along the elevation direction.

Offset paraboloidal reflectors are now commonly used in place of primary
focus-fed circularly symmetric paraboloids in order to overcome the
blockage and scattering brought about by the feed, receiver housing, and
feed-support legs.  In such a system, an elliptical section can be cut
out of a circularly symmetric paraboloid in such a way that the primary
focus is outside the main beam of the primary reflector.  It is well known
that such a system suffers a cross-polarization penalty in the form of beam
squint.  An off-axis secondary reflector can be added to the optical path
and designed to minimize the squinting at a secondary
focus.\citep{tanakam75,mizugutchay76,duanr91} The GBT and the planned
Square Kilometer Array dishes employ this design.

A circularly symmetric parabolic reflector in a Cassegrain or Gregorian
configuration can also suffer beam squint when the feed is positioned at a
secondary focus that is located off of the primary's axis of symmetry.
This arrangement obtains for multiple feeds at the Effelsberg 100-m
telescope and at the NRAO Very Large Array (VLA), where significant beam
squints have been measured.\citep{fiebigwpr91,usonc08}

If observing a large-scale region of emission for which the Stokes $I$
brightness temperature varies with position, beam squint will respond to
the first derivative of Stokes $I$ with position.  Measurements of 21-cm
Zeeman splitting can be seriously affected by spatial gradients in the
diffuse 21-cm emission interacting with the beam squint in such a way as
to produce an artificial Stokes $V$ response that exactly mimics a
Zeeman splitting
signature.\citep{trolandh82,heiles96,heilespnlbghletal01}


\subsection{Instrumental Polarization Induced by Aperture Blockage and Feed-Support Legs}
\label{rh:sec:instrumental.pol.blockage}

\begin{figure}[!tbp]
  \centering
  \includegraphics[width=\textwidth]{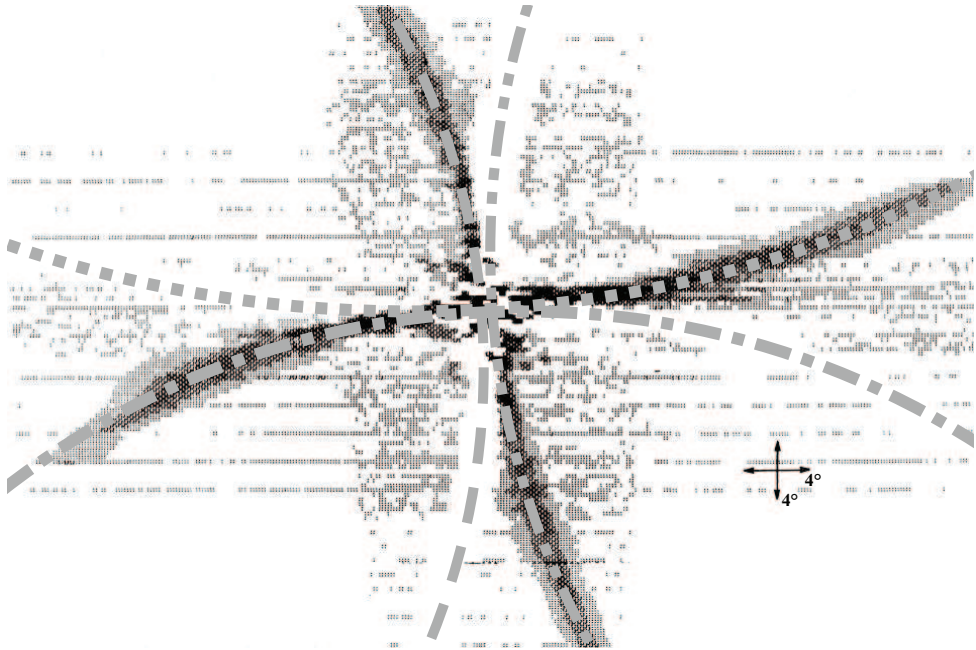}
  \begin{minipage}{\textwidth}
    \caption{Fractional Stokes $V$ sidelobe response of the now-collapsed
      85-ft Hat Creek Telescope out to $24\deg$ from beam center (adapted
      from Ref.~\refcite{trolandh82}).  Only positive Stokes $V$ is shown
      in grayscale; negative values appear as blank areas.  The four thick
      gray lines show the inner portions of the four feed-leg scatter
      rings, with each ring represented by a different line style for
      clarity. Each ring is a small circle in the sky centered on the
      direction the feed leg points, whose angular diameter is twice the
      angle the leg makes with the symmetry axis of the primary
      reflector. (Each ring---there are as many rings as there are
        feed legs---passes through and draws its energy from the main
        beam. Feed-leg scattering therefore reduces telescope gain.)  For
      each ring, the sign of the Stokes $V$ response is reversed on either
      side of beam center.}
\end{minipage}
\label{rh:fig:troland.sidelobes}
\end{figure}

Structures that block the primary aperture are also a source of polarized
sidelobes; these include feed-support legs, cables, subreflectors, and
receiver cabins.  These can produce instrumental polarization in sidelobes
both near-in to and far-out from the main beam.  While receiver cabins and
subreflectors are complex structures whose effect on the telescope's
polarized response cannot be easily modelled, the effect of feed-support
legs is relatively easy to simulate. \Fref{rh:fig:troland.sidelobes}
shows the measured Stokes $V$ response within $24\deg$ of the main beam of
the now-collapsed 85-ft Hat Creek Telescope.  The dashed lines trace four
circular features whose Stokes $V$ polarization response reverses sign on
either side of beam center.  While Refs.~\refcite{trolandh82} and \refcite{tinbergen96}
have correctly pointed out that these arcs are related to the scatter cones
generated by the quadrapod feed-support structure, they were at a loss to
explain why the circular polarization should display the observed
pattern. Modern full-polarization simulations of feed-leg scattering using
GRASP easily reveal this exact pattern, including the observed sign
reversals.\footnote{The authors haven't yet gleaned the phenomenological
  reason for the sign flip through beam center, but they take great comfort
  in seeing this empirically measured feature borne out by electromagnetic
  simulations.} Such simulations also reveal significant structure in the
Stokes $Q$ and $U$ patterns, which can affect measurements of diffuse
polarized Galactic continuum radiation.\footnote{Another significant cause
  of polarized sidelobes involves the spillover of the feed response around
  the reflector or subreflector that it illuminates; depending on the
  geometry and orientation of the telescope, the spillover sidelobe can end
  up positioned on the ground or the sky.} Because this radiation covers
the entire sky, a polarized sidelobe sitting on the sky will pick up
unpolarized radiation and alter the polarized component of the measured
signal.  Even polarized sidelobes sitting on the ground will affect the
measured on-axis polarization via two possible mechanisms: (a) the ground's
thermal radio emission is linearly polarized\citep{heilesd63}, and (b)
unpolarized off-axis Galactic emission will reflect off the ground, becoming
polarized in the process.\citep{tinbergen96,brouws76} Spectropolarimetric
studies of the 21-cm line can also be affected since the diffuse Galactic
21-cm line emission covers the entire sky: this emission can reflect off
the ground (becoming polarized in the process) and be picked up by
sidelobes sitting on the ground.

It might seem obvious that offset reflector telescopes with unblocked
apertures have no (or at least much reduced) distant sidelobes, and
therefore remove the complications just described. However, spillover is
unrelated to aperture blockage, and if an unblocked aperture is
overilluminated, producing spillover (around the primary or subreflectors),
complications remain.\footnote{Note that the GBT $L$-band feed was designed
  with too shallow a taper, such that a significant $20\deg$ diameter
  spillover sidelobe exists around the secondary with its center offset
  from the main beam by $40\deg$. At certain local sidereal times, 21-cm
  emission from the plane of the Milky Way can align with this spillover
  lobe and cause the on-axis response to change. Ref.~\refcite{robishawh09}
  showed that the instrumental polarization due to this spillover cannot be
  easily parametrized for the GBT, so that the advantages of the unblocked
  aperture are completely ruined for studies of 21-cm emission Zeeman
  splitting.}


\subsection{Putting It All Together: The Full-Stokes Off-Axis Response of
  the Arecibo Telescope}

\begin{figure}[!tbp]
  \centering
  \includegraphics[width=\textwidth]{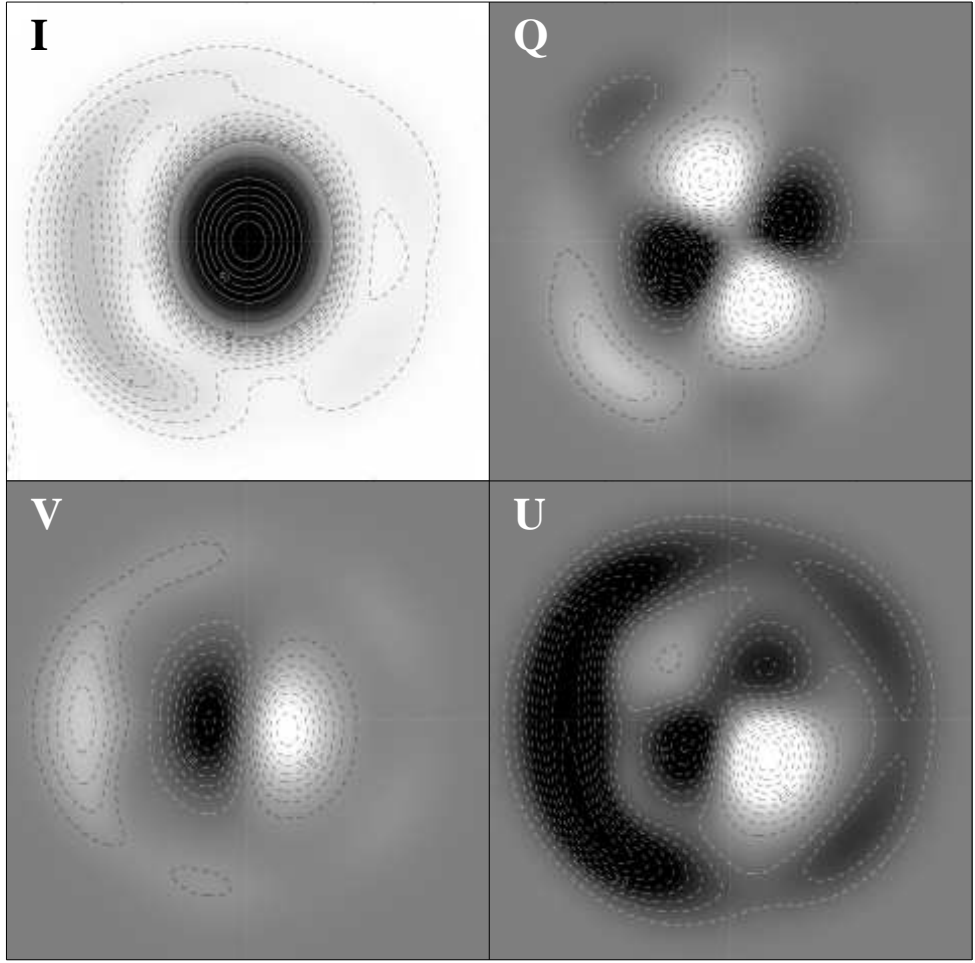}
  \caption{Arecibo beam maps at 1175 MHz for all four Stokes parameters,
    normalized to the peak main-beam Stokes $I$ response (adapted from
    Ref.~\refcite{heilespnlbghletal01}). Azimuth direction is horizontal,
    zenith angle (ZA) direction is vertical. Each map is
    19$''\times$19$''$. For Stokes $I$ \textit{(top left)} the grayscale
    covers 0---100\% of the peak Stokes $I$ (white to black), solid white
    contours cover (40\%, 50\%, ..., 90\%), solid black cover (10\%, 20\%,
    30\%), dashed black (1\%, 2\%, ..., 9\%). For Stokes $Q$ \textit{(top
      right)} and $U$ \textit{(bottom right)} the grayscale covers (white
    to black) $\pm2.8$\% of the peak Stokes $I$; thus, black is $+2.8$\%,
    white is $-2.8$\%, and gray is 0\%. For $V$ \textit{(bottom right)} the
    total range is $\pm1.6$\%.  Contours are spaced by 0.4\% for $Q$ and
    $U$, 0.2\% for $V$ (with the 0\% contour omitted for all); white
    contours are negative, black positive. The feed is native linear with
    probes at $45\deg$ with respect to the azimuth and ZA directions.}
  \label{rh:fig:beam.maps}
\end{figure}

The Arecibo telescope is a very complicated system: it has a 305-m
spherical primary reflector with shaped secondary and tertiary Gregorian
reflectors located in a focus cabin mounted on an azimuth arm. The cabin
travels along a track on the arm allowing for the beam to be pointed in
zenith angle (ZA), and the arm swings $360\deg$ in azimuth.  The azimuth
arm and focus cabin are suspended from a large multistory triangular
platform that is itself suspended via cables from three towers positioned
around the primary's perimeter.  The platform and azimuth arm block
$\sim$5--15\% of the aperture.  Despite these complexities, a team set out
to map and parametrize the polarized beam patterns of the telescope at 1175
MHz by driving the main beam across the unpolarized continuum source PKS
B1749$+$096.\citep{heiles96,heilespnlbghletal01}
\Fref{rh:fig:beam.maps} shows the resulting fractional Stokes beam
patterns; the azimuth direction is along the horizontal and the ZA
direction is along the vertical.

The far-out polarization response of this telescope---especially given the
incredibly complicated structure of the suspended platform and the shaped
subreflectors---is likely beyond the reach of accurate modelling via
software such as GRASP.  However, remarkably, some of the fundamental
instrumental polarization features that we described for a primary focus-fed
circularly symmetric paraboloid in
\sref{rh:sec:instrumental.pol.squash}--\sref{rh:sec:instrumental.pol.squint}
are clearly seen for Arecibo:\footnote{See
  Ref.~\refcite{heilespnlbghletal01} for a detailed discussion of these
  patterns.}

\begin{enumerate}

\item We saw in \sref{rh:sec:instrumental.pol.squint} that a displacement
  of the feed from the center of symmetry of the primary reflector will
  induce a beam squint in the Stokes $V$ pattern.  At Arecibo, any feed
  at the tertiary focus will always be displaced along the azimuth arm,
  which points along the ZA direction. The beam squint lobes therefore
  ought to be aligned with the azimuth direction. This is exactly what is
  seen in \fref{rh:fig:beam.maps}.

\item The expected beam squash cloverleaf pattern is seen in the Stokes $Q$
  and $U$ response. The Stokes $Q$ pattern shows the expected reversal of
  sign in the first sidelobe.  At the time these polarized beams were
  measured, the ``old'' $L$-band-wide feed was aligned with probes at
  $45\deg$ to the azimuth and ZA directions (since then, the ``new'' feed
  has replaced the ``old'' one and is aligned at $0\deg$).  In a simple
  primary focus circularly symmetric paraboloidal reflector system, the
  Stokes $Q$ squash pattern for this feed orientation would have its lobes
  aligned at $45\deg$ to the (Az,ZA) directions and the Stokes $U$ pattern
  would be aligned with (Az,ZA).  Neither is quite the case, and the $Q$
  and $U$ patterns are certainly not offset from one another by the
  expected $45\deg$.

\item The Stokes $I$ beam is highly elliptical (by design) and shows a
  significant coma lobe to the left of the main beam. The first sidelobe
  response is extreme on the coma side of the main beam and the Stokes
  $U$ pattern shows significant instrumental linear polarization response
  in this coma-side sidelobe response.

\end{enumerate}



\section{Polarization Conventions}
\label{rh:sec:conventions}

The history of polarization studies is fraught with confusion that arises
because of conventions.  As early as 1896, Pieter Zeeman, in discovering
his eponymous effect, measured the charge of (what would turn out to be)
the electron to be positive!\citet{zeeman97c} Why?  Because he had used a
mislabelled quarter-wave plate and therefore swapped his sense of
circulars.\citet{zeeman97b}

We'll say it now, and we'll say it again: When presenting polarization
results, you must \textbf{state your conventions}.


\subsection{Linear Polarization}
\label{rh:sec:linear.polarization.convention}

There are two linear polarization conventions defined by the
IAU\citet{iau74}: (1) the polarization angle $\chi$ is zero at north; and
(2) $\chi$ is measured east of north.  Thus, when represented on an image
of the sky, a line segment representing polarization rotates
counterclockwise as $\chi$ increases, and $\chi=0\deg$ corresponds to a
vertical orientation.\footnote{IAU Commissions 25 and 40 resolved to align
  the \textit{horizontal} and \textit{vertical} axes of the Stokes
  parameter reference frame along the Declination and Right Ascension axes,
  respectively.  This might seem somewhat paradoxical as we tend to think
  of Declination as the vertical equatorial axis, but the choice sensibly
  retains a right-handed coordinate system for which $\chi=0\deg$ and
  $Q/I=+1$ for completely linearly polarized radiation aligned with the
  Declination axis.}

In December 2015, the IAU sent an open letter to the astronomical community
pointing out that researchers studying the polarization of the Cosmic
Microwave Background (CMB) have been defining polarization angle to
increase clockwise on the sky.  This effectively swaps the sign of Stokes
$U$ and causes confusion for astronomers studying Galactic polarization
using CMB satellite data.


\subsection{Circular Polarization}
\label{rh:sec:conventions.circular}

If you're interested in studying circular polarization, there are a few
things you really need to worry about.\footnote{The immense confusion
  encountered in dealing with circular polarization and Stokes $V$
  definitions has been outlined at length over the last two
  decades.\citet{robishaw08,tinbergen03,hamakerb96}.} The most important
things to be aware of are:
\begin{enumerate}
\item Radio astronomers use the IEEE convention for the sense of circular
  polarization\citet{ieee69} (which has been around since 1942) and have
  been doing so at least since Pawsey \& Bracewell's 1955 seminal
  textbook\citet{pawseyb55} on the subject. Stick both your thumbs along
  the direction of propagation: whichever hand has its fingers wrapped in
  the direction that the electric field is rotating with time defines the
  handedness of the polarization sense.  To wit, if radiation is incoming,
  then stick both your thumbs towards you. If the electric field is
  rotating counterclockwise around the direction of propagation---your
  thumb---then your right hand describes the circular polarization state of
  IEEE RCP. The IEEE logo even has a drawing of the right-hand rule, in
  case you ever forget which sense is RCP.  This is \textit{opposite} to
  the definition used by physicists and optical astronomers.

\item That last point leads to a serious problem: how should astronomers
  define Stokes $V$ if optical and radio observers are using different
  definitions?  A working group chaired by Gart Westerhout tried to tackle
  this problem at the 1973 IAU meeting in Sydney\citet{iau74} by
  establishing an IAU definiton for Stokes $V$ to be IEEE RCP minus IEEE
  LCP.  Unfortunately, that definition just didn't stick---not even among
  radio astronomers.  This is likely because by 1974, the opposite
  convention was firmly established in many fundamental radio astronomy
  references.  When Cohen introduced Stokes parameters to radio astronomers
  in 1958\citep{cohen58} he had defined $V$ as IEEE $\rm{LCP}-\rm{RCP}$.
  Kraus's \emph{Radio Astronomy}\citep{kraus66}---``the bible'' for many
  generations of radio astronomers---had also defined $V$ as IEEE
  $\rm{LCP}-\rm{RCP}$ in 1966 (and again in the 1986 2nd edition).

  \hspace{4ex} Seemingly all pulsar observers (as well as Heiles and his
  Zeeman effect collaborators), unaware of the IAU definition, have used
  the Kraus $\rm{LCP}-\rm{RCP}$ definition for decades.  The pulsar crowd
  have further muddied the situation by acknowledging the discrepancy
  and---rather than adopting the IAU conventions---introducing a special
  pulsar Stokes $V$ convention that is defined oppositely from the IAU
  definition\citep{stratenmjr10}; this is implemented in their software and
  data storage definitions.

  \hspace{4ex}  We collected a sample of 53 radio Zeeman papers and found: 71\%
  failed to state whether they were using IEEE circular conventions, but we
  can give them the benefit of the doubt; 57\% failed to define their
  Stokes $V$ convention; in the cases where the Stokes $V$ convention is
  defined or can be clearly inferred, 56\% used the IAU definition.

\item The sense of circular polarization reverses upon reflection.  For
  telescopes with a feed at the primary or tertiary focus (e.g., Parkes,
  Arecibo, WSRT, GMRT), the Stokes $V$ measured by the correlator will be
  the negative of the Stokes $V$ signal incident on the primary surface.
  For telescopes with a feed at the secondary focus (e.g., $L$ band at the
  GBT, Effelsberg, VLA), the sense of Stokes $V$ measured by the correlator
  will match that of the incoming radiation.  This subtlety was overlooked
  when Verschuur\citep{verschuur68} discovered 21-cm Zeeman splitting in
  the Perseus Arm absorption feature towards Cas A using the NRAO 140-ft (a
  prime-focus telescope). He plotted Stokes $V$ as IEEE ${\rm RCP}-{\rm
    LCP}$ incident on the feed and inferred a magnetic field pointing
  towards the observer; however, in a follow-up
  publication\citep{verschuur69}, he shows the same exact Stokes $V$
  spectrum and labels it as ${\rm RCP}-{\rm LCP}$, but this time as
  incident on the dish, with a note added in proof that he had previously
  assigned an incorrect sign for the derived magnetic field vector. The
  clear lesson here is that, in addition to stating the adopted definition
  of Stokes $V$, one must state what one's Stokes $V$ spectrum
  represents---the difference in circular polarization incident on the dish
  or incident on the feed. The authors suggest that presenting Stokes $V$
  incident on the primary dish is the sensible choice: this represents the
  circular polarization state of the astronomical signal and removes the
  onus of tracking reflections from the reader.

\item The sense of circular polarization must be calibrated in order to tie
  the sign of the pseudo-Stokes correlator output ${\cal S}^{\rm cor}_{\rm
    src,3}$ to IEEE RCP or LCP.  The incoming astronomical Stokes $V$
  signal must be positive for IEEE RCP, so if an astronomical
  source\footnote{A helical antenna of known circular polarization sense
    can also be broadcast directly into the feed.}  emits a signal with net
  RCP and produces ${\cal S}^{\rm cor}_{\rm src,3}<0$, then the sign of the
  correlator output must be corrected.

\end{enumerate}


\subsection{Magnetic Field Direction}

There is a further conventional complication when comparing the {\it
  direction} of the line-of-sight component of magnetic fields in
interstellar space that have been measured by means of Zeeman splitting and
Faraday rotation.  Zeeman observers have always taken positive $\vecb{B}$
to point {\it away} from the observer, analogous to Doppler velocity, but
Manchester\citet{manchester72} changed the convention in 1972 for Faraday
rotation enthusiasts, who take positive $\vecb{B}$ to point {\it towards}
the observer in order to match with the convention that rotation measures
are positive when the field points towards the observer.

\subsection{A Factor of Two in the Stokes Parameters}

Some observatories (e.g., the VLA) define Stokes $I$ as the straight
\textit{average} of the autocorrelations in orthogonal feed responses
rather than their sum.  So if one were observing a continuum source
producing a flux density of 30 mJy in the $AA$ output and 30 mJy in the
$BB$ output, the reported Stokes $I$ value would also have a flux density
of 30 mJy.  This does not conform to the convention for the Stokes
parameters.  Stokes $I$ is defined as the \emph{sum} of the orthogonal
outputs and should have a value of 60 mJy in the above example.  The AIPS
and CASA software packages divide all the Stokes parameters by 2.  At least
they're consistent: the fractional polarization of a source should be the
same whether using the AIPS/CASA convention or the proper Stokes
convention.  But the intensities of the Stokes parameters themselves will
be half those of the proper convention, so if comparing fluxes between two
telescopes, one needs to know what conventions were used to create Stokes
$I$.  The sheer momentum of this usage means that it will never be changed,
so one must keep this in mind.\\

Given the muddled history of polarization and magnetic field conventions
over the last 50 years, there appears little chance that any single set of
conventions (even those resolved by the IAU) will be adopted by all radio
observers.  The only possible way that we can reconcile different
polarimetric observations is for you, the observer, to \textbf{state your
  conventions} when presenting results!



\section*{Acknowledgments}

We would like to thank Chat Hull, Bruce Veidt, Xuan Du, Tom Landecker, Lynn
Baker, Rick Fisher, and Phil Perillat for helpful suggestions.  It is a
pleasure, particularly for CH, to acknowledge many pleasurable years
collaborating with Prof.\ Tom Troland on polarization calibration and
measurements.  This work made extensive use of NASA's Astrophysics Data
System Abstract Service and IEEE Xplore Digital Library.



\bibliographystyle{ws-rv-van}
\bibliography{}


\end{document}